\documentclass[11pt,epsf,amssymb,ulem,qsymbols]{article}

\usepackage{jheppub}
\usepackage{graphicx}
\usepackage{bm}
\usepackage{bbm}

\newcommand{\bea}{\begin{eqnarray}}
\newcommand{\eea}{\end{eqnarray}}
\newcommand{\beq}{\begin{equation}}
\newcommand{\eeq}{\end{equation}}
\newcommand{\bit}{\begin{itemize}}   
\newcommand{\eit}{\end{itemize}}

\def\acal{{\cal A}}
\def\ccal{{\cal C}}
\def\dcal{{\cal D}}

\def\gcal{{\cal G}}
\def\hcal{{\cal H}}
\def\lcal{{\cal L}}
\def\ncal{{\cal N}}
\def\ocal{{\cal O}}

\def\zBB{{\mathbbm Z}}

\def\alpbf{{\bm\alpha}}
\def\betbf{{\bm\beta}}	

\def\pibf{{\bm\pi}}
\def\xibf{{\bm\xi}}
\def\psibf{{\bm\psi}}

\def\kk{{\bf k}}
\def\lll{{\bf l}}
\def\pp{{\bf p}}
\def\qq{{\bf q}}
\def\uu{{\bf u}}
\def\xx{{\bf x}}
\def\EE{{\bf E}}

\def\KK{{\bf K}}
\def\TT{{\bf T}}
\def\VV{{\bf V}}

\def\half{\frac12}
\def\im{{\bf Im}}
\def\re{{\bf Re}}
\def\ui{U(1)}
\def\then{{\quad\Rightarrow\quad}}

\def\up#1{^{\left( #1 \right) }}
\def\inv#1{\frac1{#1}}
\def\vevof#1{\left\langle #1 \right\rangle}
\def\tr#1{{\rm tr}\left\{ #1 \right\}}
\def\su#1{{SU(#1)}}
\def\deriva#1#2#3{\left(\frac{\partial #1}{\partial #2}\right)_{#3}}

\def\tev{\hbox{TeV}}
\def\gev{\hbox{GeV}}
\def\mev{\hbox{MeV}}
\def\sm{Standard Model}
\def\lhs{left hand side\ }

\def\hc{{\rm H.c.}}
\def\gdm{\gcal_{\rm DM}}
\def\hdm{\hcal_{\rm DM}}
\def\ne{n\up{{\rm eq}}}

\def\tn{\tilde{n}}
\def\lh{\lambda_h} 
\def\lv{\lambda_V} 
\def\mx{\kappa}
\def\Yeq{Y^{{\rm(eq)}}}

\def\vt{\vartheta_{SM}}
\def\vtt{\vartheta_V}

\def\dqq#1{\deriva{#1}{q^2}{q=0}}

\title{\boldmath Pionic Dark Matter}

\author[a,1]{Subhaditya Bhattacharya,\note{Corresponding author.}}
\author[a,b]{ Bla\v{z}enka Meli\'c,}
\author[a]{Jos\'e Wudka}

\affiliation[a]{Department of Physics {\it\&} Astronomy,
University of California  Riverside\\ Riverside CA 92521-0413, USA}
\affiliation[b]{Rudjer Bo\v{s}kovi\'c Institute, Theoretical Physics Division\\
P.O.Box 180, HR-10002 Zagreb, Croatia.}
\emailAdd{subhaditya.bhattacharya@ucr.edu}
\emailAdd{blazenka.melic@irb.hr}
\emailAdd{jose.wudka@ucr.edu}

\abstract{
We study a phenomenological model where the lightest
dark matter (DM) particles are the
pseudo-Goldstone excitations associated with a 
spontaneously broken symmetry, and transforming
linearly with respect to an unbroken group $\hdm $.
For definiteness we take $\hdm = \su N$ and
assume the Goldstone particles are
bosons; in parallel
with QCD, we refer to these particles as dark-matter pions.
This scenario is in contrast to the common assumption
that DM fields transform linearly under the full symmetry
of the model. 
We illustrate the formalism by treating in detail the case of $\hdm=\su2$,
in particular
 we calculate all the interactions relevant for the Boltzmann equations,
which we solve numerically; we also derive approximate analytic solutions 
and show their consistency with the numerical results.
We then compare the results with the constraints
derived from the cold DM and direct detection experiments and derive the corresponding restrictions
on the model parameters.}


\begin{document} 
\maketitle
\flushbottom


\section{\label{sec-0}Introduction}

Dark matter (DM) is the most promising hypothesis proposed to 
explain astrophysical and cosmological observations 
related to the motion of stars in galaxies \cite{Oort:1932}, the motion
of galaxies in clusters \cite{Zwicky1,Zwicky2,peebles}, structure formation \cite{Klypin:1992sf}
and the inhomogeneities in the CMBR \cite{WMAP,PLANCK}. Having
not direct experimental information about this  
component of the universe the theoretical efforts
to understand DM have been couched within realistic
extensions of the Standard model (SM)~\cite{susy1,susy2,ed1,ed2,ed3,lh1,scotino}, or have taken
a purely phenomenological approach~\cite{zurek1,zurek2,zurek3,silk,eg1,eh1}, in which case 
simplicity has been used as a guide and constraint.

In this publication we will investigate a phenomenological
model for DM based on general assumptions concerning the
dark sector, explicitly, we will assume that the lightest
particles in that sector are the pseudo-Goldstone bosons 
resulting from a broken symmetry~\cite{review}. Operationally
this implies that the lightest particles (that we take as scalars
for simplicity) transform non-linearly under a
continuous symmetry group, a situation similar to 
the one occurring in low energy hadron physics. Accordingly, 
we will refer to them as dark matter pions (DMP)
(we emphasize however, that these are quite distinct form the
pions in the hadronic sector, in particular
they do {\em not} have direct couplings to the standard model
(SM)  $W^\pm$ and 
photon (in this we fundamentally differ from the assumptions made in \cite{vecchi}). 
This approach is in contrast with most phenomenological approaches 
where the dark-sector fields are assumed to transform under
a discrete symmetry, or linearly under a continuous one~\cite{zurek1,zurek2,zurek3,silk}.

In the following we will study this type of DM model 
 based on the nonlinear realization of a spontaneously 
broken symmetry group $\gdm $. 
However, given the difficulties of hot dark matter gas in dealing with 
structure formation \cite{hot-DM}, we will also assume that the Goldstone bosons 
receive their masses through an explicit breaking of the
original symmetry. We also require that all SM particles are
singlets under the dark-sector symmetries and that
the dark particles are singlets under the SM local
symmetries. 

The interaction between these two sectors (SM and DM)
is presumably effected by the exchange of some
heavy mediators whose nature we do not need
to specify, but only assume are much heavier than the typical scales in either
sector. Therefore the typical interactions are of the
form
\beq
\lcal_{\rm DM-SM} \sim \inv{\Lambda^n} \ocal_{\rm DM}
 \ocal_{\rm SM}\,,
\label{eq:generic.int}
\eeq
where $\ocal_{DM},\ocal_{SM}$ are operators invariant under the internal symmetries
of the corresponding sector, but they need not be Lorentz invariant
(though, of course, $\lcal_{\rm DM-SM}$ must be). The details of these interaacitons 
will be elaborated blow.

This paper is organized as follows: in the next section we describe the
formalism behind our model, and construct the Lagrangian we will use in our
calculations. In section \ref{sec:DMP.int} we calculate the SM-DM interactions
that we then use in sections \ref{sec:thermal.history} and \ref{sec:su2.BE}
to derive the relic abundance of this type of dark matter. These results
are compared with the experimental constraints in section \ref{sec:exp.const} 
with our brief conclusions are presented in section \ref{sec:concl}. A few details
are relegated to the two appendices.

\section{Nonlinear realization of $\gdm$}

Models where the symmetry is nonlinearly realized 
have been extensively studied (see, e.g. \cite{Coleman:1969sm}); here we
summarize some of the results for completeness.
We assume there is a subgroup $ \hdm \subset \gdm$ under which
the vacuum is invariant and, following \cite{Coleman:1969sm}, 
we denote the generators of $\hdm$ by $V_i$ and the
remaining generators of $ \gdm$ by $T_a $. 
Then the fields can be chosen as $\{\pibf,\psibf\} $ with the following properties:
\bit
\item Under $\hdm$ they transform linearly:
$ \pibf \to \dcal(h) \pibf,~ \psibf \to D(h) \psibf $ for $ h \in \hdm$; where
$\dcal$ and $D$ are some matrix representations of $ \hdm $.

\item Under a general $ g \in \gdm $
\beq
\pibf \to\xibf(\pibf,g)\,, \quad \psibf \to D\left(e^{\uu.\VV}\right) \psibf\,;~
\uu = \uu(\pibf,g)\,,
\eeq
where $D$ is the same representation as above, and $\xibf$ and \uu\ are defined by
\beq
g e^{\pibf.\TT} = e^{\xibf.\TT} e^{\uu.\VV} \,.
\eeq
\eit
Note that the transformation of $\pibf$ depends only on $g$ and $ \pibf $,
and is non-linear;
while that of $ \psibf$ depends on $g$, $\psibf$ and $ \pibf$.
Because of their transformation properties the $ \pibf$ are
massless and correspond to the Goldstone bosons generated under
the spontaneous breaking $ \gdm \to \hdm $, and accordingly
the number of these fields equals that of the broken generators
$ T_a $. We will refer to the $ \pibf$ as the ``dark-matter pions'' (DMP)
or dark pions.

To be specific we concentrate on the familiar case~\cite{chiral.reference,gasser.leutwyler} of a unitary
chiral theory where $ \gdm = \su N \times \su N $
and $\hdm = \su N$, the diagonal subgroup. In this case the
above general formalism is realized by introducing a 
unitary field $\Sigma$ and transforms as
\beq
\Sigma \to L \Sigma R^\dagger \qquad L,\,R \in \su N \,,
\eeq
where $ \Sigma = \exp( i \pibf.\TT/f)$
and  $f$ is a mass scale associated with the 
spontaneous breaking of the symmetry. The diagonal
subgroup corresponds to the choice $ R = L $.

As it is well known~\cite{gasser.leutwyler,georgi},
the leading fully chirally invariant operator is
\beq
\lcal\up0 = f^2 \tr{ \partial_\mu \Sigma^\dagger \, \partial^\mu \Sigma} \,.
\label{eq:dmp.l0}
\eeq
Expanding (\ref{eq:dmp.l0}) in terms of the $\pibf $ we find 
that this Lagrangian describes a
series of massless particles\footnote{We will
not be concerned here with coherent excitations
that might be stabilized by higher-derivative
operators that describe dark baryons \cite{skyrme}.} which
are difficult (though not impossible \cite{Archidiacono:2013cha})
 to reconcile with 
structure formation. We will therefore also
include an explicit breaking of the $ \gdm$
symmetry that generate a mass for these excitations;
for the chiral model this corresponds to a term of the form
\beq
\lcal_{\rm mass} = \half f^2 \left( M^2 \tr\Sigma + {\rm H.c.} \right) \,.
\label{eq:dmp.lmass}
\eeq
This term is invariant under the diagonal (unbroken) subgroup $\hdm$.

In order to construct the DM-SM interactions of
the form (\ref{eq:generic.int}) we need the 
list of the lowest-dimensional SM gauge-invariant
(tough not necessarily Lorentz invariant) operators.
These are easily listed; for dimension $\le2$ we have
\beq
{\rm dim}~2: ~ |\phi|^2 \,, B_{\mu\nu} \,,
\eeq
where $ \phi$ denotes the SM scalar doublet and
$B$ the hypercharge gauge field.
The dimension 3 operators (that we will not use here) are
$ \phi^\dagger D_\mu\phi$ and $\bar\psi \gamma_\mu \psi'$,
where $ \psi $ and $ \psi' $ are any two fermion fields
carrying the same gauge group representation 
(e.g. $e_R$ and $ \tau_R$); higher dimensional operators are similarly
constructed.

Then, the simplest DM-SM coupling is clearly
\beq
\lcal_{\Sigma-\phi} = \half \lh \left( |\phi|^2 - v^2 \right)
\tr{ \partial_\mu \Sigma^\dagger \, \partial^\mu \Sigma}\,,
\label{eq:dmp.lSf}
\eeq
where $v = \vevof\phi  \sim 174 \,\gev $.

The coupling $ \Sigma $ to $ B_{\mu\nu} $ is less straightforward
since there are no $\gdm$-invariant
operators that can be constructed out of $ \Sigma $ and
its derivatives and which transforms as
the $ (0,1) + (1,0)$ representation of the Lorentz 
group\footnote{Those terms become available for models with
two chiral fields $ \Sigma_{1,2} $ that  transform in the same way.}.
Noting however, that (\ref{eq:dmp.lmass}) is invariant only under the 
diagonal subgroup $ \hdm $, we will only require the $ \Sigma-B$ coupling to 
have the same property, and in this case,
\beq
\lcal_{\Sigma-B} = B^{\mu\nu} \left(\lv \tr{ \Sigma^\dagger \partial_\mu \Sigma \partial_\nu \Sigma^\dagger } + \hc \right) \,.
\label{eq:dmp.lSB}
\eeq

For our choices of $\gdm $ and $ \hdm $ the Lagrangian for our model
is obtained from (\ref{eq:dmp.l0}, \ref{eq:dmp.lmass}, \ref{eq:dmp.lSf}, \ref{eq:dmp.lSB});
explicitly,
\bea
\lcal &=& 
\half \left[ f^2  + \lh \left( |\phi|^2 - v^2
\right) \right] \tr{\partial_\mu \Sigma^\dagger \, \partial^\mu \Sigma} \cr
&& \quad + \half  f^2 \left( M^2 \tr\Sigma + \hc \right)
+  B^{\mu\nu} \left(\lv \tr{ \Sigma^\dagger \partial_\mu \Sigma \partial_\nu \Sigma^\dagger } + \hc \right) \,,
\label{eq:dmp.ltot}
\eea
where, as before,
\beq
\Sigma = \exp\left( \frac if \pi_a T_a \right) \,.
\eeq
In parallel with the usual strong-interaction
pions, we will call $f$ the DMP decay constant. 

The
$T_a$ are the broken Hermitian generators normalized by
\beq
\tr{T_a T_b} = \delta_{ab}\,,
\label{eq:T.norm.1}
\eeq
and obeying
\beq
[ T_a , T_b] = i f_{abc} T_c 
\label{eq:T.norm.2}
\eeq
(with $ a,b, \ldots = 1, 2, \ldots, N^2-1 $). In the Cartan basis
with root generators $ T_{\pm\alpbf} $ and Cartan generators $T_i $
we have \cite{group_theory}
\beq
[T_i, T_j] =0 \,,\quad
[T_i, T_\alpbf] = \alpha_i T_\alpbf \,, \quad
[T_\alpbf, T_\betbf] = N_{\alpbf,\betbf} T_{\alpbf+\betbf}\,,
\eeq
where $N_{\alpbf,\betbf}=0 $ if $\alpbf+\betbf $ is not a root.

We could also add another $\phi-\pi$ coupling by replacing
\beq
M^2 \to M^2(\phi) =  M^2 + \lv' \left(|\phi|^2 - v^2 \right) \,.
\eeq
To lowest order this coupling is of the form $ |\phi|^2 \pibf^2 $ and its
effects have been studied extensively \cite{higgs-portal}. Given our interest
in studying the effects of the new interactions listed
in (\ref{eq:dmp.ltot}) we will neglect $ \lv'$ in the following.

Writing $ \Sigma = \exp(i \sigma) $ and using
\beq
\delta \Sigma = i \int_0^1 du \, e^{i(1-u)\sigma} 
\delta\sigma \; e^{i u \sigma}\,, \quad \sigma = \pibf.\TT/f
\eeq
the Lagrangian can be written (in a Hermitian basis)
\bea
\lcal &=& \half \left( 1 + \lh \frac{|\phi|^2 - v^2}{f^2}
\right) \partial_\mu \pi_a \partial^\mu \pi_b \,  g_{ab} 
+ \inv2 M^2 f^2 \tr{ \Sigma + \Sigma^\dagger} \cr  && \quad
 - \inv{f^2} B^{\mu\nu} \partial_\mu \pi_a \, \partial_\nu\pi_b \,
g_{ac} f_{cbd} \, \im\left( \lv \tr{T_d \Sigma^\dagger} \right) \cr&&\cr
&=& \half (\partial \pibf )^2 -
\half M^2 \pibf^2  
+  \frac{\lh v}{\sqrt{2}f^2} h  (\partial \pibf )^2 
+ \frac{\lh }{4 f^2} h^2 (\partial \pibf )^2 
- \frac{\re(\lv)}{f^3}
B^{\mu\nu} f_{abc} \partial_\mu \pi_a \, \partial_\nu\pi_b \, \pi_c
+ \cdots  \,, 
\cr && 
\label{eq:ltot2}
\eea
where
\beq
g_{ab} = \int_{-1}^1 du\, (1-|u|) \tr{ e^{i u \sigma} \,
T_a \, e^{-i u \sigma} T_b} \,,
\eeq
and $h$ is the Higgs field; in unitary (SM) gauge $ \phi^T =  (v + h/\sqrt{2}) (0,1)$.

In the Cartan basis, 
\bea
\pibf^2 &=& \sum_i \pi_i^2 + \sum_\alpbf | \pi_\alpbf|^2 \,, \quad 
\pi_{-\alpbf} = \pi^\dagger_\alpbf \,,\cr
&&\cr
(\partial\pibf)^2 &=& \sum_i (\partial \pi_i)^2 + \sum_\alpbf | \partial \pi_\alpbf|^2
=\sum_i (\partial \pi_i)^2 +2 \sum_{\alpbf>0} | \partial 	\pi_\alpbf|^2 \,, \cr
&&\cr
B^{\mu\nu} f_{abc} \partial_\mu \pi_a \, \partial_\nu\pi_b \, \pi_c 
&=& i B^{\mu\nu} \Biggl[ \sum_{\alpbf,\betbf }N_{\alpbf,\betbf}\pi_{\alpbf+\betbf}^\dagger \,
\partial_\mu\pi_\alpbf \, \partial_\nu \pi_\betbf 
\cr &&  \qquad \quad+\ 
\sum_{i,\alpbf} \alpha_i\,
\partial_\nu\pi_\alpbf^\dagger \left( 2 \pi_{\alpbf} \partial_\mu \pi_i
- \pi_i \partial_\mu \pi_{\alpbf} \right) \Biggr]\,,
\eea
and for the case of $ N=2$ (that we will develop later as a
specific illustrative case):
\bea
\pibf^2 &=&  \pi_{o}^2 + 2\pi_+ \pi_- \,,\cr
&&\cr
(\partial\pibf)^2 &=& (\partial \pi_o)^2 + 2 \partial \pi_+
\partial \pi_- \,,\cr
&&\cr
B^{\mu\nu} f_{abc} \partial_\mu \pi_a \, \partial_\nu\pi_b \, \pi_c &=&
-2i B^{\mu\nu} \left[ (\partial_\mu \pi_o) ( \pi_- {\stackrel\leftrightarrow
\partial} \pi_+) + \pi_o \, \partial_\mu\pi_+
\partial_\nu \pi_- \right] \,.
\eea
where $\pi_o $ is associated with the $\su2$ Cartan generator,
and $ \pi_\pm = \pi_{\pm\alpbf } $, where $\alpbf $ is the
single root in this group.

\subsection{Conserved currents}
\label{sec:cons.chg.}

The Lagrangian (\ref{eq:dmp.ltot}) is invariant under
the global transformations
\beq
\Sigma \to V^\dagger \Sigma V \,; \quad V \in \su N\,,
\eeq
which give rise to a set of conserved Noetherian currents
\bea
J^\mu_b &=&  \left( 1 + \lh \frac{|\phi|^2 - v^2}{f^2}
\right) \partial^\mu \pi_d g_{ad} \pi_c f_{bca} 
 - \frac2{f^2} B^{\mu\nu} \, f_{bca}\pi_c g_{ae} f_{edf} \partial_\nu\pi_d 
 \im\left( \lv \tr{T^f \Sigma^\dagger} \right) \,.
\cr &&
\eea

Ignoring the interactions with the SM 
the canonical momentum  are
$ \wp_a = g_{ab} \dot \pi_a $ in terms
of which
the charges (again ignoring the SM interactions) become
\beq
Q_b = \int d^3\xx  J^0_b = \int d^3\xx  \;\pi_c f_{bca} \wp_a
\eeq
and (ignoring possible sigma terms and other anomalies~\cite{Jackiw:1986dr}) 
satisfy the algebra
\beq
[Q_a,Q_b] = i f_{abc} Q_c\,,
\eeq
as expected.

The number of commuting conserved charges equals the rank of the group,
which, in a Cartan basis, can be conveniently chosen as those associated with the
$ \pi_i $:
\beq
 [Q_i , Q_j] =0 ; \qquad\
Q_i = \sum_\alpbf \alpha_i \int d^3\xx  \pi_\alpbf \wp_\alpbf \,.
\label{eq:qi}
\eeq 
Assuming that these relations  do not exhibit 
commutator anomalies~\cite{Jackiw:1986dr}  the charges $Q_i$ will be conserved; in particular
this property will be reflected in the Boltzmann equations.
 It follows from the expression
for $Q_i$ that the $ \pi_i$ carry no charge, while $\pi_{\pm\alpbf} $
carry opposite $i$-charges when $ \alpha_i \not =0 $.

\subsection{Parameters of the model}
\label{sec:parameters}

The model we consider has then 4 parameters: the DMP mass $M$, the DMP
 decay constant $f$, the coupling constant of the DMP to
the Higgs $\lh$, and $\lv$, the  coupling constant of the DMP to
the hypercharge vector field $B$ (from which follow the coupling to the
 $Z$ boson and the photon). 

In  the calculations below we will take $\lv $ coupling to be real
with magnitude
\beq
\lv =  0.63\,.
\label{lambda'}
\eeq
We will see later that as far as the Boltzmann equations are concerned,
any change in $ \lv$ can be absorbed in a redefinition of the
other parameters (cf. the end of Sec. \ref{sec:su2.BE}), 
so this choice does not represent a loss of
generality and is made for computational ease only.
It is worth noting that according to naive dimensional analysis (NDA)~\cite{Manohar:1983md}
its value is $ \lv \sim g'/(4\pi)^2 \simeq 0.0023 $, where
$g'$ is the $\ui_Y$ gauge coupling constant in the \sm.

For the rest of the parameters we impose just some loose
constraints. We require that  
\begin{equation}
\lh < 1
\end{equation}
in order to ensure the model remain perturbative\footnote{In
imposing this constraint we are being conservative as the 
perturbative unitarity limit is in fact $ \lh < 4 \pi $.}.
We will see later that all the experimental constraints on the
model also have simple scaling dependence on the couplings $ \lh $
(see Sec.\ref{sec:CDM.constraints}),
so this constraint will also not restrict the generality of
our results.

Since we assume that the DMP are the pseudo-Goldstone
bosons of some underlying theory and are generated by the breaking
of $ \gdm $ to $ \hdm $ at some scale $ \Lambda $, consistency
of the resulting chiral model requires~\cite{georgi}
\begin{equation}
4 \pi f \gg M\,.
\label{eq:f-M}
\end{equation}
For large values of $N$ the \lhs\ is expected to be 
suppressed by a factor of $ 1/\sqrt{N}$~\cite{Chivukula:1992nw}, which we do not include because we will
restrict ourselves to low values of $N$.

Another constraint can be derived
by requiring loop corrections not to dominate over the tree-level terms.
In particular this should hold for the radiative corrections generated by the term 
proportional to $\lv $ in (\ref{eq:ltot2}),
which includes vertices of the form 
$ (\lv/f^{n+2}) Z_{\mu\nu} \partial^\mu\pi \, \partial^\nu \pi
\, \pi^n $. Two such vertices will generate loop corrections
to the $ \partial^\mu\pi \, \partial_\mu \pi\, \pi^k/f^k $
vertex of the first term in (\ref{eq:ltot2}):
\beq
\vbox{\vfil\hbox to 1.9in{\vspace{-.5in}\includegraphics[height=2.5cm]{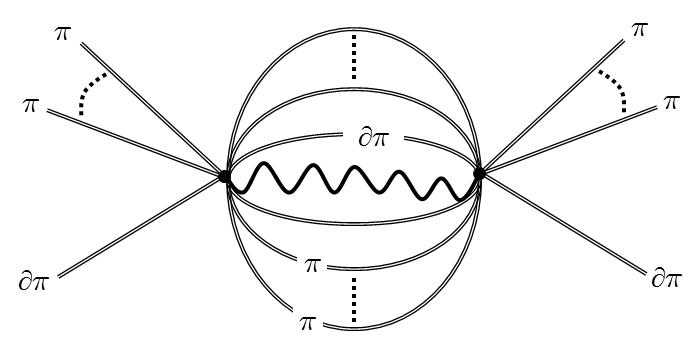}}\vfil}
\sim \frac{( 4 \pi \lv )^2}{f^k} \left( \frac M{4\pi f} \right)^{2L+2} \qquad (L=\rm{number~of~loops})
\vspace{.3in}
\label{eq:loops}
\eeq
where we have assumed that all the terms in (\ref{eq:ltot2}) that explicitly
violate $ \gdm $ are associated with the scale $M$, which we have used as an UV cutoff.
We require (\ref{eq:loops}) not to be larger than the tree-level
contribution, which implies (since $L$ can be arbitrarily large)\footnote{This can be refined by introducing 
the loop symmetry factor of $1/\Gamma(L)$; the lower bound
on $ 4 \pi f/M$ in terms of $ x = 4 \pi \lv$ then becomes: $\sqrt{x}$ for $ x > 1 $;
$ x^{1/3} $ for $ 1\ge x\ge 1/8 $, and below $x=0.125 $ it is well approximated by
$ - (1/\ln x) + [3 \ln(-\ln x)- \ln ( 2\pi)]/[2 (\ln x)^2] $. We will not, however, use
these more complicated relations below.}
\beq
f \ge \left[ {\rm max}\{ 4 \pi \lv\,,\, 1 \} \right]^{1/2} \frac M{4\pi} \,.
\label{eq:pert.constraint}
\eeq

\section{DMP interactions \label{sec:DMP.int}}

In this section we calculate the cross sections for the
processes that dominate the Boltzmann equations that
describe possible equilibration between the dark and SM sectors, 
and within the dark sector. The
relevant  interactions (\ref{eq:ltot2}) separate into
those that involve only DMP, and those that involve DMP and the
SM scalar $ \phi $ or the vector boson $B$.
We also derive the reactions
relevant for direct detection of the DMP. 
In all the calculations below we only consider $ 2 \to 2 $ 
processes and will use the Cartan basis for the DMP.

\subsection{DMP $ \to $ SM interactions}

\begin{figure}[thb]
\centering
\centerline{\includegraphics[height=2cm]{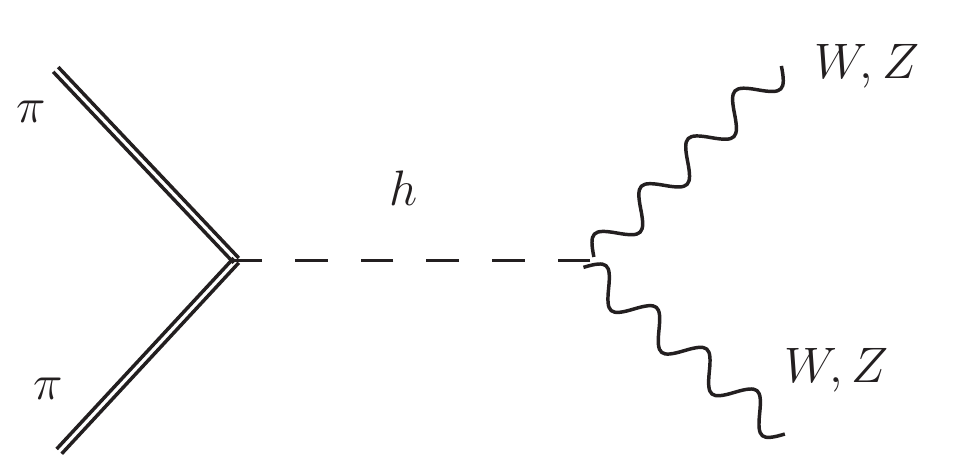}
\hskip 20 pt \includegraphics[height=2cm]{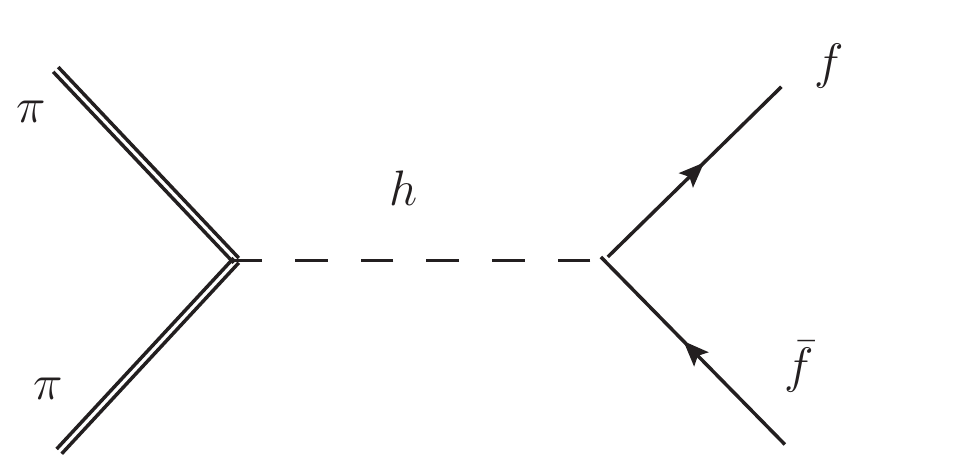}}
\vskip 30 pt
\includegraphics[height=2.5cm]{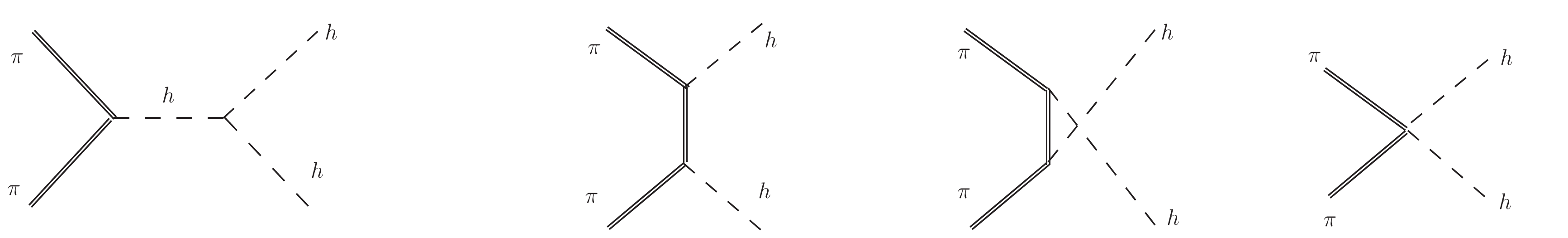}
\caption{DMP $\to$ SM particle diagrams.} 
\label{DMP_SM}
\end{figure}

There are two kinds of reactions:

\paragraph{Processes with only SM particles in the final state.} These are of the form
\bea
\pi_i \pi_i \to h^* \to {\rm SM}\,, &\quad&
\pi_i \pi_i \to hh \,,
\cr
\pi_\alpbf \pi_{-\alpbf} \to h^* \to {\rm SM} \,,&\quad&
\pi_\alpbf \pi_{-\alpbf} \to hh\,,
\eea
for which the interaction terms in (\ref{eq:ltot2}) are
\beq
\lcal_{h-2\pi} =  \left(
\frac{ v \lh}{\sqrt{2} \,f^2} h
+
\frac{\lh}{4f^2} h^2
\right)
\left[ \sum_i (\partial\pi_i)^2 + 2\sum_{\alpbf>0} | \partial \pi_\alpbf|^2 \right]\,,
\label{eq:h.pi.pi}
\eeq
and the processes are shown in Fig.\ref{DMP_SM}.

The cross sections for these processes are:
\bea
\sigma(\pi \pi \to W^+W^-) &=&  
\frac{12\mx_W^2 -4 \mx_W + 1}{4\mx_\pi^2} \beta_W \, \sigma_{SM} \,,\cr
&& \cr
\sigma(\pi \pi \to  ZZ) &=&  
\frac{12\mx_Z^2 -4 \mx_Z + 1}{8 \mx_\pi^2} \beta_Z \, \sigma_{SM} \,,\cr
&& \cr
\sigma(\pi \pi \to  f \bar f) &=& \frac{ \mx_f}{2 \mx_\pi^2} \beta_f^3 \sigma_{SM} \,,\cr
&& \cr
\sigma(\pi \pi \to  hh) &=& \frac{s\lh^2}{1024\pi f^4} 
\frac{\beta_h}\beta\,
\Biggl\{ \left[ \frac{(1-2\mx_\pi)(1+\mx_h)}{1-\mx_h}  - \frac{4 \lh v^2}{f^2}(1-8\mx_\pi+2\mx_h) \right]^2
\cr&& \quad
- \frac{4 \lh v^2}{f^2} \left[ \frac{(1-2\mx_\pi)(1+\mx_h)}{1-\mx_h}  
       - \frac{4 \lh v^2}{f^2} \frac{1-8 \mx_\pi+4 \mx_\pi^2+3 \mx_h(4  \mx_\pi- \mx_h)}{1-2\mx_h} \right]\Upsilon 
\cr&& \quad
+ \frac{16\lh^2 v^4}{f^4} \Biggl[
\frac{2(\mx_h-2\mx_\pi)^4}{\mx_\pi+\mx_h(\mx_h-4\mx_\pi)}\Biggr]
\Biggr\}  \,,
\label{eq:dmp-sm}
\eea
where 
\bea
 \mx_i = m_i^2/s\,, ~ (m_\pi = M)  \,; &\qquad& \beta_i = \sqrt{1-4 \mx_i}\,; \cr
&&\cr
 \sigma_{SM} = \frac{s\lh^2 }{16\pi f^4}\frac{\mx_\pi^2}{\beta_\pi}\,
\frac{(1-2\mx_\pi)^2}{(1-\mx_h)^2 + \mx_h (\Gamma_h^2/s)} \,; &\quad&
\Upsilon = \frac{4(\mx_h-2\mx_\pi)^2}{\beta_\pi \beta_h}
\ln \left( \frac{1-2 \mx_h + \beta_\pi \beta_h}{1-2 \mx_h -\beta_\pi \beta_h}
\right)\,.
\nonumber \\
\label{eq:mp-sm.defs}
\eea
We neglected the Higgs width in the expression for $ \sigma( \pi\pi \to hh) $
since it is never resonant (resonance occurs at $ s \sim m_h^2$ while
the reaction occurs only if $ s > 4 m_h^2 $) and
current data \cite{higgs} suggests $ \Gamma_h \simeq \Gamma_h\up{{\rm SM}} \simeq 4$ MeV and
$ m_h =125$ GeV so that $ \Gamma_h\up{{\rm SM}}/m_h \simeq 3.2\times 10^{-6} $. 
For the $W$, $Z$ and $t$ reactions we can also ignore
$\Gamma_h$ in $\sigma_{SM}$ (defined in eq. \ref{eq:mp-sm.defs}); 
the same is true for the other reactions if $M > m_h/2 $. 

\paragraph{Processes involving DMP in the final state.}

\begin{figure}[thb]
\centering
\centerline{\includegraphics[height=2cm]{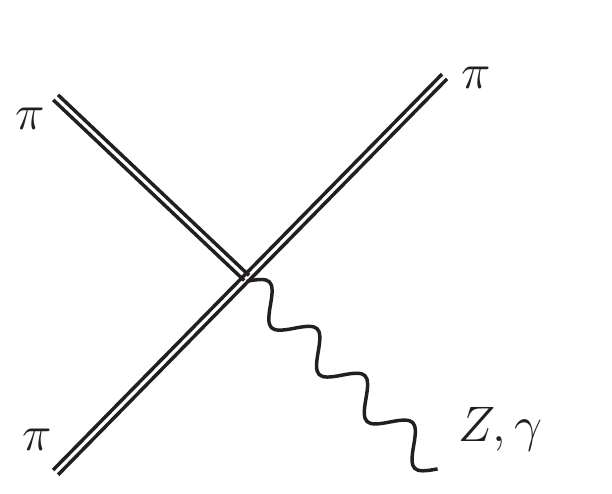}
\hskip 20 pt \includegraphics[height=2cm]{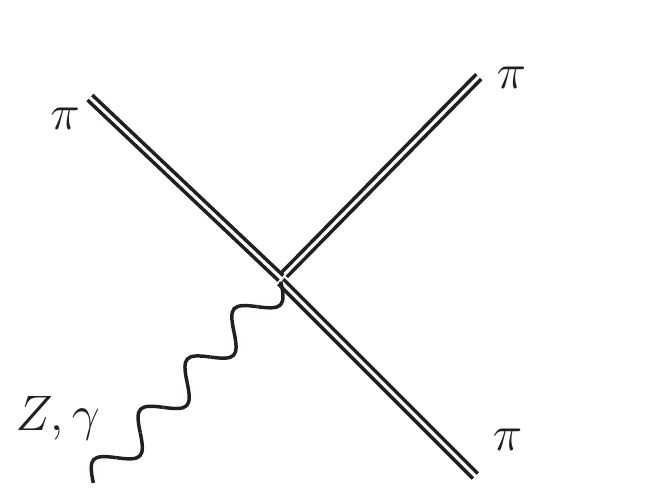}}
\caption{DMP scattering with $Z$ and $\gamma$.} 
\label{DMP_Zgamma}
\end{figure}

These correspond to $ \pi \pi \leftrightarrow \pi Z/\gamma  $ for which the
Lagrangian is given by
\bea
\lcal_{Z-3\pi} &=&
\frac{i \lv}{f^3} 
i B^{\mu\nu} \left\{
\sum_{\alpbf,\betbf}
\partial_\nu\pi_\alpbf  \partial_\mu\pi^\dagger_\betbf 
N_{\alpbf , -\betbf} \pi_{\alpbf - \betbf}^\dagger 
\right. \cr && \quad\qquad \left.
+ \sum_\alpbf (\partial_\nu \pi_\alpbf) \left[
2 \pi_\alpbf^\dagger (\partial_\mu \alpbf.\pibf)
- 2 (\alpbf.\pibf)(\partial_\mu \pi_\alpbf^\dagger)  \right]
\right\}\,.
\eea
So there are 3 types of reactions
(the first present only for $ \su N,~ N>2$):
\beq
\begin{array}{ccccc}
\pi_\alpbf(p) &\pi_\betbf(q) &\leftrightarrow& \pi_{\alpbf + \betbf}(l)& V(k) \,, \cr
\pi_\alpbf(p) &\pi_{-\alpbf}(q) &\leftrightarrow& \pi_i(l)& V(k) \,,\cr
\pi_\alpbf(p) & \pi_i(q) &\leftrightarrow&  \pi_\alpbf(l)& V (k) \,
\end{array}
\eeq
($V$ denotes $Z$ or $\gamma$), which are
presented in Fig.\ref{DMP_Zgamma}.
The cross sections are
\bea
\sigma(\pi_\alpbf \pi_\alpbf^\dagger \to \pi_i \, V )
= \sigma(\pi_\alpbf \pi_i \to  \pi_\alpbf \, V ) =
\alpha_i^2 \frac{K_V}P \sigma^V \,, &&
\sigma(\pi_\alpbf \pi_\betbf \to  \pi_{\alpbf+\betbf} \,V ) =
|N_{\alpbf,\betbf}|^2  \frac{K_V}P \sigma^{V} \,,\cr
&&\cr
\sigma(\pi_i \,V \to \pi_\alpbf \pi_\alpbf^\dagger )
= \sigma(\pi_\alpbf \,V \to \pi_\alpbf \pi_i )=
\frac{\alpha_i^2}{s_V}  \frac P{K_V} \sigma^{V} \,,&&
\sigma(\pi_{\alpbf+\betbf} \, V \to \pi_\alpbf \pi_\betbf ) =
\frac{|N_{\alpbf,\betbf}|^2}{s_V}  \frac P{K_V} \sigma^V \,,
\cr &&
\eea
where $ s_V$ the number
of spin degrees of freedom: $s_Z=3,\; s_\gamma=2 $, and 
\bea
\sigma^Z &=& \left( \frac{3 s_{\rm w} \lv}{f^3} \right)^2 \frac{P^2}{16\pi s}
\left[ \left(s - M^2 - \inv3 m_Z^2 \right)^2 - 
\frac43\left(s- \frac49 m_Z^2\right)K_Z^2\right] \,,\cr 
\sigma^\gamma &=& \left( \frac{3 c_{\rm w} \lv}{f^3} \right)^2 
\frac{P^2}{24\pi s} \left(s - M^2 \right)^2 \,.
\label{eq:sigmaV}
\eea
In the center of momentum (CM) frame  $K_V= |\kk| = |\lll|$ denotes the magnitude of the $V$ 
3-momentum,  and $P= |\pp| = | \qq| $ the magnitude of
the 3-momentum of the
pions not paired with the vector boson:
\beq
K_V^2 = \frac{\lambda(s,m_V^2,M^2)}{4s} \,, \qquad
P^2 = \frac{\lambda(s,M^2,M^2)}{4s} \,,
\label{eq:P.KV}
\eeq
with
\beq
\lambda(a,b,c) = a^2 + b^2 + c^2 - 2 a b - 2 b c - 2 c a \,.
\label{eq:def.of.lambda}
\eeq

\subsection{Direct-detection reaction}
\label{sec:direct.detection}

The most important process that can contribute to the scattering
of the DMP off heavy nuclei (relevant for direct DM detection~\cite{DAMA,CDMS,Xenon})
is $ \pi \psi \to \pi \psi $, where $\psi$ is SM fermion, and occurs
through a $t$-channel $h$ exchange. 
The averaged amplitude-squared is
\beq
\overline{|\acal|^2} = \left( \frac{m_\psi \lh}{2 f^2} \right)^2 \left( 
\frac{t-2M^2}{t-m_h^2} \right)^2 (4 m_\psi^2 -t )\,,
\eeq
so that, in the CM frame, the corresponding cross section for this process is given by 
\bea
\sigma ( \pi \psi \to \pi \psi ) &=& 
\inv{16\pi s} \left( \frac{m_\psi \lh}{2 f^2} \right)^2 \Biggl\{
2(P^2-m_h^2+2M^2+2 m_\psi^2) 
- \frac{(m_h^2 - 4 m_\psi^2)(m_h^2-2M^2)^2}{m_h^2(m_h^2 + 4 P^2)}
\cr && \qquad \qquad \qquad
+ \frac{(2M^2+8 m_\psi^2-3m_h^2)(2M^2-m_h^2) }{4P^2}
\ln \left| \frac{4P^2+m_h^2}{m_h^2} \right|
\Biggr\} \,,
\eea
where $P$ denotes the momentum of the incoming particles in the CM frame.
When $ M,m_h \gg P, m_f $ this cross section is approximated by 
\beq
\sigma(\pi \psi \to \pi\psi)
\simeq
\inv{4\pi s} \left( \frac{ m_\psi \lh M^2}{m_h^2 f^2} \right)^2 \left(m_\psi^2 + \frac{P^2}{2} \right) 
\quad (M,m_h \gg P, m_f).
\label{eq:direct.detection}
\eeq

At low momentum transfer the effective interaction obtained from
integrating the Higgs using (\ref{eq:h.pi.pi}) and the \sm\
$h \bar f f $ interaction $-(m_\psi/v) h \bar\psi\psi $ is 
\beq
\lcal\up{\rm eff}_{\pi\pi\psi\psi} = 
-\left(\frac{ \sqrt{2} m_\psi \lh M^2}{m_h^2\,f^2} \right)
\half \pibf^2 \, \bar\psi\psi \,.
\label{eq:leff.psi.pi}
\eeq

\subsection{Pure DMP scattering}

\begin{figure}[thb]
\centering
\centerline{\includegraphics[height=2cm]{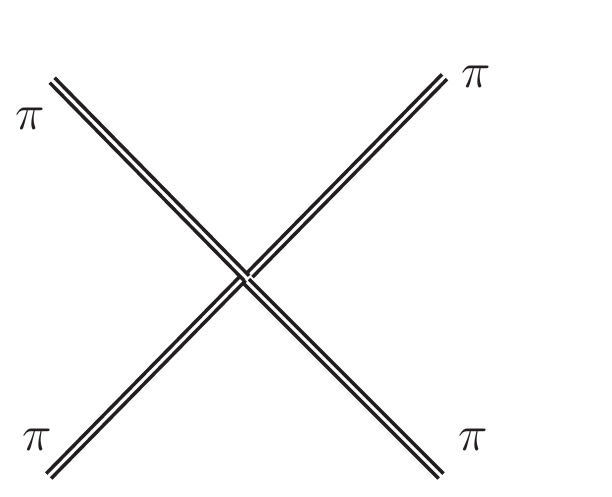}}
\caption{DMP $\to$ DMP scattering diagram.} 
\label{DMP_DMP}
\end{figure}

Finally, we obtain the cross sections responsible for equilibrium within the DMP
sector, $ \pi \pi \to \pi \pi $, Fig.\ref{DMP_DMP}. The lowest-order terms (taking $M$ real) 
in (\ref{eq:ltot2}) are
\beq
\lcal = \half ( \partial \pibf)^2 - \half M^2 \pibf^2 + 
 \frac N{16 f^2(N^2-2)} \left[ (\partial \pibf^2)^2 - \mu^2 (\pibf^2)^2
\right] \,, 
\label{eq:l4pi.mu}
\eeq
where 
\beq
 \mu^2 = \frac{6 N^2 -4}{N^2(N^2+1)}M^2
\eeq
 and we have dropped terms that vanish on shell and will no contribute to the S-matrix.

In terms of DMP defined in the Cartan basis
\beq
\pibf^2 = \sum_i \pi_i^2 + 2 \sum_{\alpbf > 0} \pi_\alpbf \pi_{-\alpbf} \,;
\qquad \pi_\alpbf^\dagger = \pi_{-\alpbf} \,,
\eeq
we have the following reactions: 
\beq
\begin{array}{llll}
{\rm reaction} & {\rm \qquad Lagrangian} & {\rm amplitude} & {\rm cross section} 
\cr \hline
ii \to jj ~ (i \not= j) & -(u/4) \pi_i^2 ( \square + \mu^2) \pi_j^2 
& i u (s-\mu^2) & \sigma_0/2  \cr
ii \to ii  & -(u/8) \pi_i^2 ( \square + \mu^2) \pi_i^2 
& i u (4 M^2 - 3\mu^2) & u^2 \left( M^2- \frac34\mu^2 \right)^2/(2\pi s)  \cr
ii \to \alpbf \bar\alpbf &  -(u/2)  \pi_i^2 ( \square + \mu^2)|\pi_\alpbf|^2 
& i u (s-\mu^2) & \sigma_0  \cr
\alpbf \bar\alpbf \to ii &  -(u/2)  \pi_i^2 ( \square + \mu^2)|\pi_\alpbf|^2 
& i u (s-\mu^2) & \sigma_0/2  \cr
\alpbf\bar\alpbf \to \betbf\bar\betbf ~ (\alpbf\not= \betbf)
&  -u |\pi_\betbf|^2 ( \square + \mu^2)|\pi_\alpbf|^2 
& i u (s-\mu^2) & \sigma_0 \cr
\alpbf\bar\alpbf \to \alpbf\bar\alpbf
&  -(u/2) |\pi_\alpbf|^2 ( \square + \mu^2)|\pi_\alpbf|^2 
& 2i u (M^2-\mu^2) & u^2 \left( M^2- \mu^2 \right)^2/(4\pi s) 
\end{array}
\label{eq:p4.scattering}
\eeq
where $ \bar\alpbf = - \alpbf, ~ \bar\betbf = - \betbf$, and
\beq
\sigma_0 = \frac{u^2(s-\mu^2)^2}{16 \pi s} \, , \qquad  u = \frac N{2 f^2 (N^2-2)} \,.
\label{eq:sigma0.u}
\eeq

\subsection{Decays of SM particles to DMP}

Limits on the DMP parameters can be derived
either from collider reactions or 
from potential deviations from SM decays. 
Reactions of the form $ f\bar f \to \pi \pi $,
where $f$ is a SM fermion,
or $W$ fusion reactions $ WW \to \pi \pi $,
would mimic neutrino production at colliders. The
limits, however, are very weak since these 
processes proceed through a virtual $h$ and so the
amplitude will be proportional to  small Yukawa coupling,
or, for the case of heavy initial quarks,
suppressed distribution functions.

\begin{figure}[thb]
\centering
\centerline{\includegraphics[height=2cm]{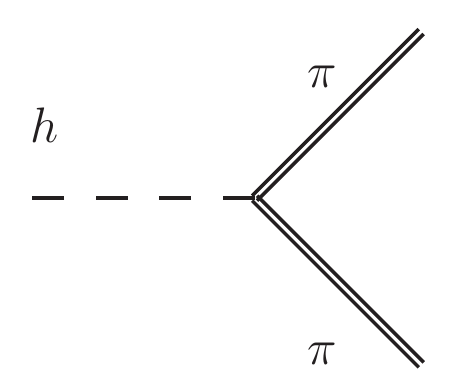}
\hskip 40 pt \includegraphics[height=2cm]{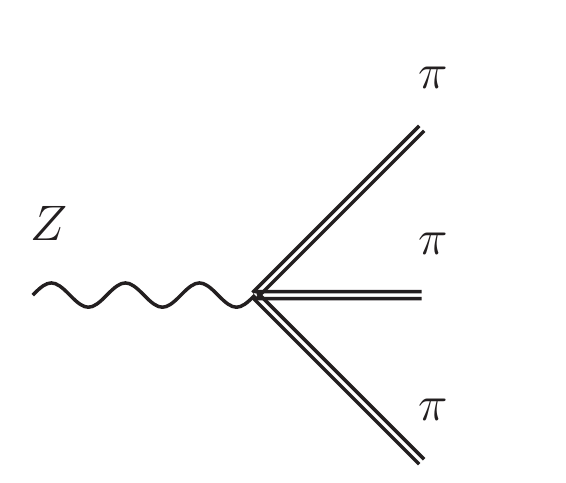}}
\caption{SM particle decays to DMP.} 
\label{DMP_dec}
\end{figure}

The main limits are then derived form the two
leading decay processes, Fig.\ref{DMP_dec}, namely,
$ h \to \pi \pi $ and $ Z \to \pi \pi \pi$, to which
we now turn.

\paragraph{$h\to \pibf\pibf$ decay}
\label{sec:hpp}

Using (\ref{eq:ltot2}) and choosing  a Hermitian $\pi $ basis
we find that the width is given by
\beq
\Gamma(h\to \pi_a \pi_b)  = \Gamma_{h\pi\pi}\delta_{ab}\,;
\quad \Gamma_{h\pi\pi}=
\frac{(\lh v)^2}{16\pi m_h}
\left( \frac{m_h^2 -2M^2}{2f^2}\right)^2
\sqrt{1 - \frac{4 M^2}{m_h^2}} \, \theta ( m_h - 2 M) \,;
\label{eq:hppwidth}
\eeq
in the Cartan basis
$ \Gamma( h \to \pi_i\pi_i ) = \Gamma_{h\pi\pi} $ and
$ \Gamma( h \to \pi_\alpbf\pi_{-\alpbf})  = 2\Gamma_{h\pi\pi} $.
Recent data \cite{higgs} favors a Higgs decay close to the SM prediction
of $ \sim 4 \,\mev $ and a mass $ m_h \sim  125 \,\gev $; 
this requires $ M > m_h/2 $, or $ M< m_h/2$ and $ 
\Gamma_{h\pi\pi} < 4 \,\mev $, hence the constraint we use is
\beq
f > 5.9  |\lh|^{1/2} |7812.5 - M^2|^{1/2}
\left[ 1 - \left(\frac M{62.5} \right)^2 \right]^{1/8} \,,
\qquad M < 62.5 ~ (M~\hbox{in}~\gev)\,.
\label{eq:higgs.constraint}
\eeq

In the numerical solutions to Boltzmann equations for DMP for 
the $\su2$ case (discussed below),
we consider DMP masses in the interval $50 \,\gev \le M \le 2000 \,\gev$ so the 
$h\to \pibf\pibf$ constraint plays an important role only for comparatively
small values of $M$.

\paragraph{ $Z\to \pibf\pibf\pibf$ decay}

The calculation is straightforward; using again a Hermitian DMP basis we find
\beq
\Gamma(Z \to \pi \pi \pi) = 
\frac{M^7 s_{\rm w}^2 \lv^2}{15 \, (8\pi f^2)^3 \, r^{5/2}}
\left( \sum_{a > b > c} |f_{abc}|^2 \right)
\left[ p_E \EE(c) + p_K \KK(c) \right]\,,
\label{eq:Zpppwidth}
\eeq
where $s_{\rm w}= \sin \theta_{\rm w}$, while 
$\EE,\,\KK$ denote the usual Elliptic functions, and
\bea
p_E &=& (3 r^8+ 394 r^6 - 720 r^4 + 54 r^2-243)\,, \cr
p_K &=& - \half(r-1 )^3 (20 r^6 + 63 r^5 + 99 r^4 + 522 r^3 + 918 r^2 + 567 r+ 243 )\,, \cr
c &=& -\frac{(r-3) (1 + r)^3}{16 r} \,,\cr
r &=& \frac{m_Z}M\,.
\eea
For $\hdm  = \su N$ and our normalization conventions
(\ref{eq:T.norm.1},\ref{eq:T.norm.2})
the summation involving the structure constants is given by
\beq
\sum_{a>b>c} |f_{abc}|^2 = \inv{3!} \sum_{a,b,c} |f_{abc}|^2 = \frac{ N(N^2-1)}3\,.
\eeq

Using the uncertainty in the invisible width of the $Z$, $\Gamma(Z)_{\rm inv} $ we have limit
\beq
{\Gamma(Z\to\pi\pi\pi)}{} <  3 \times 10^{-3}\Gamma(Z)_{\rm inv} =
3 \times 10^{-3}  \frac{g^2 m_Z}{32 \pi c_{\rm w}^2}\,,
\eeq
which implies
\beq
55.4 >
N(N^2-1)
\left( \frac{ m_Z^3 \lv }{f^3} \right)^2 Q \,,
\label{eq:z3pi}
\eeq
where
\beq
Q = r^{-19/2}[ p_E \EE(u) + p_K \KK(u)]\,.
\eeq
The function  $Q$ is monotonic; it vanishes  as $ r \to 3 $ 
and approaches $0.75$  as $ r \to \infty $. Taking $N=2$,
and $ \lv = 0.63 $, the most conservative
limit (corresponding to taking $Q=3/4 $) corresponds to
\beq
f > 51.43 \,\gev \quad  \left (M < \frac{m_z}3 \right )\,.
\eeq
When  $ \lv = 0.063 $, this limit becomes $f > 23.87 \,\gev $.

In the numerical analysis, we choose to work with DMP mass $\ge 50 \,\gev$ and therefore the constraint from
$Z\to \pi \pi \pi $ is of no importance.

\section{Thermal history of DMP}
\label{sec:thermal.history}

We now turn to the derivation of the relic abundance of DMP.
We follow the standard treatment (see e.g. \cite{Kolb:1990vq}) and
will consider only $2 \to 2 $ processes.

\subsection{Boltzmann equations}

The change in the number density of particle of type $a$ 
due to collisions and the expansion of the universe is
given by
\bea
\dot n_a + 3 H n_a &=& - \ccal_a \,, \cr
\ccal_a &=&\sum_{b,c,d} \int d\Phi
 |\acal_{a+b \to c+d}|^2(f_a f_b - f_c f_d) \,,\cr
d\Phi &=&  d\Pi_a \, d\Pi_b \, d\Pi_c \, d\Pi_d
(2\pi)^4 \delta\up4(p_a + p_b - p_c - p_d) \,,
\label{eq:boltzmann}
\eea
where $ d\Pi $ denotes the phase-space volume 
\beq
d\Pi = \frac g{2E_\pp}\frac{d^3 \pp}{(2\pi)^3} \,,
\eeq
and $g$ is the number of internal degrees of freedom.
The amplitude-squared
$|\acal|^2 $ for
the $a + b \to c + d $ process is understood
to be
{\em averaged} over initial
and final states, and to include symmetry factors for identical 
particles in the final states. The functions  
$f$ are the particle phase-space distribution functions;
the corresponding particle
number density is
\beq
n = g \int \frac{d^3\pp}{(2\pi)^3} f \,.
\eeq

We will assume that interactions are such that
kinetic equilibrium is maintained~\cite{dodelson}; we will also assume that 
particles densities are sufficiently small
to ignore the effects of quantum statistics. In this case the energy dependence in the distribution functions
is given by the Boltzmann factor: $ f = \zeta \exp(- E/T )$.
Since we are interested in the epoch when the DMP first
decouple, all distribution functions will have the same temperature $T$;
this will continue after decoupling provided no mass thresholds are crossed,
or phase transitions occur. 

The equilibrium distributions for a particle of mass $m$ is given by
\bea
\ne  &=& g z \int \frac{d^3\pp}{(2\pi)^3}e^{-E/T} 
=\frac{z g m^3}{2\pi^2} \frac{K_2(x)}x\,, \quad
E=\sqrt{m^2 + \pp^2}\,,~x= \frac mT
\label{eq:neq.def}
\eea
where $z$ is the fugacity in equilibrium. For the SM $z_{\rm SM} = 1$ to
very good accuracy~\cite{weinberg-old}; for the DMP, however, we will allow non-zero
chemical potentials. Using the definition in (\ref{eq:qi}) and the discussion
below it, it follows that 
\beq
\mu\up i_j =0 \,,\qquad \mu\up i_\alpbf = - \mu\up i_{-\alpbf} \,,
\label{eq:chem.pots}
\eeq
where $ \mu_a\up i $ denotes the chemical potential
for particle $a$ associated with charge $Q_i $ so that $ z\not=1$ for those particles with
non-zero conserved charges, as defined in Sec. \ref{sec:cons.chg.}.

Substituting these definitions in the expression for $ \ccal $ and
using the standard definition of the scattering cross section $ \sigma $ we
find
\bea
\ccal_a &=&
 \sum_{b,c,d}\left( \tn_a \tn_b - \frac{\ne_a \ne_b}{\ne_c \ne_d} \tn_c \tn_d \right) \vevof{\sigma v}_{a+b\to c+d} \,,\cr 
&& \cr
\vevof{\sigma v}_{a+b\to c+d}
&=&\frac{ T g_a g_b}{2\,(2\pi)^4\ne_a \ne_b} 
\int_{s_o}^\infty ds\, \frac{\lambda(s, m_b^2, m_a^2)}{\sqrt{s}} 
K_1(\sqrt{s}/T ) \, \sigma_{a+b \to c+ d}(s) \,,
\label{eq:def.of.ccal}
\eea
where $\tilde{n} = z n$, $ s = ( p_a + p_b)^2 = m_a^2 + m_b^2 + 2 p_a.p_b$, $ \lambda(a,b,c)$ is defined in (\ref{eq:def.of.lambda}), 
and 
\beq
s_o = {\rm max}\{(m_a+m_b)^2, (m_c+m_d)^2 \}\,.
\eeq 
In the definition of $s_o$ we used the condition (contained in the cross section)
that $s$ should be large enough to create $c$ and $d$. 

For the pure DMP scattering processes that appear in the Boltzmann equations
the averaged cross sections can be evaluated
in closed form.  We obtain, for example 
\bea
\vevof{\sigma v}_{\pi_i\pi_i\to \pi_\alpbf \pi_{-\alpbf}} &=& \inv{8 } \frac x{z_{\pi_a}z_{\pi_b}[K_2(x)]^2} \inv{M^5}
\int_{4 M^2}^\infty ds\, \sqrt{s}\,(s-4 M^2)
K_1(\sqrt{s}/T ) \, \sigma_0 \cr
&=& \frac{4u^2M^2}{\pi } \inv{z_{\pi_a}z_{\pi_b} x^3[K_2(x)]^2}
\left[ \frac{ B x^2+3}x K_2(2x) + \frac{B^2 x^2 + 6}4 K_1(2x)\right]\,, 
\label{eq:sv.4p}
\eea
with similar expressions for the other relevant processes; in deriving this
we used (\ref{eq:p4.scattering}) and (\ref{eq:sigma0.u}).
For the relevant initial states ($\pi_i \pi_i $ or $ \pi_\alpbf \pi_{-\alpbf}$) we have
$ z_i = 1 = z_{\alpbf} z_{-\alpbf} $ so that in all cases of interest
(see below)  we can replace
$ z_{\pi_a}z_{\pi_b} \to 1 $.
Also $u$ is defined in (\ref{eq:sigma0.u}), while $B$ is  defined as
\beq
B = 1 -\frac{\mu^2}{4M^2} = \frac{N^4 -\half N^2 +1 }{N^2(N^2+1)}
\eeq
and $ \mu $ is given in (\ref{eq:l4pi.mu}). In deriving the above result
we used
\beq
\int_1^\infty dy (y^2-1)^n K_1(2 x y) = \frac{n!}2 \frac{K_n(2x)}{x^{n+1}} \,.
\eeq

With the above preliminaries we can now find  the relevant collision terms
$ \ccal_a$ (c.f Eq.(\ref{eq:boltzmann})) for the cases $ a = \pi_i $
and $ a = \pi_\alpbf $
that we abbreviate as $ \ccal_i$ and $ \ccal_{\cal_\alpbf} $ respectively.
We will assume that all 
SM particles remain in equilibrium,
so that $ n_{\rm SM} = \ne_{\rm SM}$. The tables of the relevant reactions  
(which do not cancel in $\ccal_{i,\alpha}$ ) are
\bea
&\begin{array}{|c|c|}
\hline
\multicolumn{2}{|c|}{a =\pi_\alpbf} \cr\hline\hline
b & c/d \cr\hline
\pi_{-\alpbf} & W^+W^-,\; ZZ,\; f \bar f,\; hh,\; \pi_j V,\; \pi_j\pi_j,\; \pi_\betbf\pi_{-\betbf}\cr \hline
V & \pi_j\pi_\alpbf,\; \pi_\betbf\pi_{\alpbf-\betbf} \cr\hline 
\pi_j & V \pi_\alpbf \cr\hline
\pi_\betbf & V \pi_{\alpbf+ \betbf} \cr\hline
\end{array}& \cr
&&\vbox{\bigskip}\cr
 &\begin{array}{|c|c|}
\hline
\multicolumn{2}{|c|}{a =\pi_i} \cr\hline\hline
b & c/d \cr\hline
\pi_i & W^+W^-,\; ZZ,\; f\bar f,\; hh,\; \pi_j\pi_j,\; \pi_\betbf \pi_{-\betbf} \cr\hline
\pi_\betbf & V \pi_\betbf \cr\hline
\end{array} &
\eea
where $V$ represents $Z$ or $ \gamma$,
$\betbf \not=-\alpbf $, and a summation over $j$ and $\betbf$ is assumed. 

Now, using (\ref{eq:def.of.ccal}) and noting that (\ref{eq:chem.pots})
implies
\beq
\tn_i = n_i \,,\qquad
\tn_\alpbf \tn_{-\alpbf}= n_\alpbf n_{-\alpbf} \,,
\eeq
and similarly for the equilibrium densities, we find
\bea
\ccal_i &=& (n_i^2 - \ne_i{}^2) \vevof{\sigma v}_{\pi_i\pi_i\to SM} 
+ \sum_\alpbf \tn_\alpbf \left( n_i  - \ne_i \right) \left[
\vevof{\sigma v}_{\pi_i \pi_\alpbf \to \gamma \pi_\alpbf} 
+ \vevof{\sigma v}_{\pi_i \pi_\alpbf \to Z \pi_\alpbf} \right] \cr && \quad
+ \sum_\alpbf \ne_i\left( \frac{n_i}{\ne_i} - \frac{n_\alpbf}{\ne_\alpbf} 
\frac{n_{-\alpbf}}{\ne_{-\alpbf}} \right)
\left [ \ne_Z \vevof{\sigma v}_{\pi_i Z \to \pi_\alpbf \pi_{-\alpbf}} + 
 \ne_\gamma \vevof{\sigma v}_{\pi_i \gamma \to \pi_\alpbf \pi_{-\alpbf}} \right ]
\cr && \quad
+ \sum_{j\not=i}\left(n_i^2 - \frac{\ne_i{}^2}{\ne_j{}^2}n_j^2 \right) 
\vevof{\sigma v}_{\pi_i\pi_i\to \pi_j\pi_j}  
+\sum_\alpbf\left (n_i^2 - \frac{\ne_i{}^2}{\ne_\alpbf \ne_{-\alpbf}}
n_\alpbf n_{-\alpbf} \right) 
\vevof{\sigma v}_{\pi_i\pi_i\to \pi_\alpbf\pi_{-\alpbf}}
\cr && 
\label{eq:Ci}
\eea
and
\bea
\ccal_\alpbf &=& (n_\alpbf n_{-\alpbf} - \ne_\alpbf \ne_{-\alpbf}) 
\vevof{\sigma v}_{\pi_\alpbf\pi_{-\alpbf}\to SM} \cr && \quad
+ \sum_i \left( n_\alpbf n_{-\alpbf} - \ne_\alpbf \ne_{-\alpbf}
\frac{n_i}{\ne_i} \right) 
\left[
\vevof{\sigma v}_{\pi_\alpbf \pi_{-\alpbf} \to \pi_i \gamma}
+
\vevof{\sigma v}_{\pi_\alpbf \pi_{-\alpbf} \to \pi_i Z} \right]
\cr && \quad
+ \sum_i \left(n_\alpbf n_{-\alpbf} - 
\frac{\ne_\alpbf \ne_{-\alpbf}}{\ne_i{}^2} n_i^2 \right) 
\vevof{\sigma v}_{\pi_\alpbf\pi_{-\alpbf}\to \pi_i\pi_i} \cr &&\quad 
+ \sum_{\betbf\not=\pm\alpbf}\left(n_\alpbf n_{-\alpbf} - 
\frac{\ne_\alpbf \ne_{-\alpbf}}{\ne_\betbf \ne_{-\betbf}}
n_\betbf n_{-\betbf} \right) 
\vevof{\sigma v}_{\pi_\alpbf\pi_{-\alpbf}\to \pi_\betbf\pi_{-\betbf}}\,,
\label{eq:Ca}
\eea
where the contributions coming from $\pi_\alpbf V \to \pi _i \pi_\alpbf$ 
($V=Z,\,\gamma$) and
$\pi_\alpbf \pi_i \to  V  \pi_\alpbf$ cancel, as do those from $\pi_\alpbf \pi_\betbf \to V \pi_{\alpbf + \betbf}$ 
and $\pi_\alpbf V \to \pi_\betbf \pi_{\alpbf- \betbf}$. We have also defined, using (\ref{eq:def.of.ccal}),
\bea
\vevof{\sigma v}_{\pi_i\pi_i\to SM} &=& 
\vevof{\sigma v}_{\pi_i\pi_i\to WW} +
\vevof{\sigma v}_{\pi_i\pi_i\to ZZ} +
\sum_f \vevof{\sigma v}_{\pi_i\pi_i\to ff} +
\vevof{\sigma v}_{\pi_i\pi_i\to hh} \cr 
&=& \frac{ T }{32 \pi^4 \ne_i{}^2} 
\int_0^\infty ds\, s^{3/2} \, K_1(\sqrt{s}/T) \beta^2 \biggl[
\sigma_{\pi_i\pi_i\to WW}
+ \sigma_{\pi_i\pi_i\to ZZ} \cr && \qquad
+\sum_f  \sigma_{\pi_i\pi_i\to f \bar f} 
+ \sigma_{\pi_i\pi_i\to hh} \biggr] 
\eea
and similarly for $ \pi_\alpbf \pi_{-\alpbf} \to SM$.

\subsection{Contributions from SM$\to$DMP decays}

The effects of Higgs decays into DMP, when kinematically allowed, 
can be included in the Boltzmann equation
in two equivalent ways. We can include them in the
total $h$  width:
\beq
\Gamma_h = \Gamma_h^{\rm SM} + \Gamma(h\to \pi \pi) 
\eeq
and use this expression in the cross sections involving Higgs exchange.
Or, alternatively, we can exclude these effects from the Higgs propagators 
(see e.g. \cite{Frigerio:2011in}) :
\bea
\frac{m_h \Gamma_h}{(s - m_h^2)^2 + m_h^2\Gamma_h^2}  &\to& 
\frac{m_h \Gamma_h}{(s - m_h^2)^2 + m_h^2\Gamma_h^2}  - \pi \delta(s - m_h^2)\Theta (s - 4 m_i^2)\,.
\eea
and include them in suitable additions
$\ccal_{i,\alpha}\up{\rm decay}$ to the collision terms; explicitly (see Appendix \ref{sec:app.a})
\bea
\ccal_i\up{\rm decay} &=& N_H^i \ne_H \frac{K_1(x_H)}{K_2(x_H)} \Gamma(h \to \pi\pi) 
\label{eq:CZ}
\eea
where $N_h\up i $ counts the number of produced $\pi_i$:
$N_h\up i=2!$ for $ h \to \pi_i \pi_i $ and $N_h\up i=1$ for $ h \to \pi_\alpbf \pi_{-\alpbf}$; 
$\Gamma(h \to \pi\pi)$ is given in 
(\ref{eq:hppwidth}), 
and $x_i = m_i/T$. An analogous equation holds for 
$\ccal_\alpbf\up{\rm decay}$. 

If we assume that the recently observed particle at the LHC \cite{higgs}
is the SM Higgs, it's very small total width ensures that the effects
from Higgs decay to DMP are negligible.
We have checked that for realistic DMP masses the contribution of $Z\to\pi\pi\pi$ decays 
in the Boltzmann equations (see Appendix \ref{sec:app.a}) are also negligible. 

\section{Solving the Boltzmann equations for the $\su2$ case}
\label{sec:su2.BE}

The simplest non-trivial group is $ \hdm = \su2$, which we consider as an illustrative
example of the formalism; the same approach can be used for any $N$, though with
the calculations become increasingly cumbersome. For $N=2$ there is a single conserved 
charge and 3 DMP states that we label as $o,\pm$,
with the first associated with the Cartan generator. 

As usually we find it convenient to rewrite the Boltzmann equations (BE) (\ref{eq:boltzmann}, \ref{eq:Ci}, \ref{eq:Ca}) 
by defining
\beq
x = \frac  MT \,, \qquad Y_r = \inv s n_r \,, \qquad \Yeq_r = \inv s \ne_r \,,
\eeq
where $T$ denotes the photon temperature and $s$  the entropy density:
\beq
s = \frac{2 \pi^2}{45} g_s (T) T^3 \,; \quad
g_s(T) = \sum_k r_k g_k \left (\frac{T_k}{T} \right )^3\theta (T-m_k) \,;
\eeq
here $k$ runs over all particles, $T_k$ is the temperature of particle $k$
and $g_k$ its number of internal degrees of freedom, and
$ r_k=1\,(7/8)$ when $k$ is a bosons (fermion).
We will also make use of  Friedman's equation,
\beq
H^2 = \frac{8\pi G}{3} \rho = \frac{4\pi^3 G}{45} g(T) \,T^4 \,;
\quad g(T) = \sum_k r_k g_k \left (\frac{T_k}{T} \right )^4\theta (T-m_k) \,.
\eeq
In the following we will take $T_k$
for all SM particles (assuming $T$ is above that of the $ e^+ e^-$
annihilation epoch), so that $g_s(T) = g(T)$; we use the expression for
$g(T)$ in Ref. \cite{Laine}. The explicit form
of the equilibrium distribution is
\beq
\Yeq_r = \frac{45}{4\pi^4} \frac{g_r}{g_s(T)} z_r x_r^2 K_2(x_r)
\stackrel{x_r \gg1}\longrightarrow a_r z_r x_r^{3/2} e^{-x_r}\,;
\qquad
x_r = \frac{m_r}T\,,~ ~ a_r= \frac{45}{4\pi^4} \sqrt{\frac\pi2}\frac{g_r}{g_s(T)}
\label{eq:Yeq.def}
\eeq
where $z_r $ is the fugacity for  particle $r$ and $ g_r $ the number
of internal degrees of freedom.

We will also consider model parameters where the SM and DM sectors are in 
equilibrium for temperatures $ T > T_f$, such that $ T_f > M $, so that the
region of interest is $ x > 1 $ and the
DMP will not contribute\footnote{
If $ T_f > M $ then the situation is more complicated, the DMP maintain
a temperature $T_\pi $ which is initially $T_f $, but then is determined by $ s_\pi(T_\pi) R^3
= s_\pi(T_f) R_d^3 $ and is in general different form the photon temperature.}
to the effective number of relativistic
degrees of freedom $ g(T) = g_{SM}(T)$.

In terms of $Y$ the Boltzmann equations take the form
\bea
\frac{d\,Y_r}{d\,x} = - \sqrt{\frac{\pi g(T)}{45 G}}\frac{M}{x^2}  C_r (Y)
\;, \qquad C_r(Y) = \inv{s^2} \ccal_r \,, \qquad (r=o,\pm)
\label{eq:dY}
\eea
where the collision terms are 
\bea
C_o(Y) 
&=&  \left({Y_o^2} - { \Yeq_o}^2 \right) \vevof{\sigma v}_{\pi_o\pi_o\to SM} 
+\left (Y_o^2 -  Y_+ Y_- \right) \vevof{\sigma v}_{\pi_o\pi_o\to \pi_+\pi_{-}}  
\cr &&~
+ \left[ Y_o \Yeq_o - Y_+ Y_- + (Y_+ +Y_-)(Y_o -  \Yeq_o)  
\right] \vevof{\sigma v}_{\pi_+ \pi_- \to \pi_o V} \,,
\eea
and
\bea
C_\pm(Y) 
&=& \left( {Y_+ Y_- } - { \Yeq_o}^2 \right) 
 \vevof{\sigma v}_{\pi_o\pi_o\to SM}  
 + \left(  Y_+ Y_- -  Y_o^2 \right) \vevof{\sigma v}_{\pi_+\pi_{-}\to \pi_o\pi_o} \cr &&~
+ \left( {Y_+ Y_- } - { Y_o \Yeq_o} \right) 
 \vevof{\sigma v}_{\pi_+ \pi_{-} \to \pi_o V}  \,,
\eea
where we used $\Yeq_+ \Yeq_- = {\Yeq_o}^2$, and
also 
\beq
\Yeq_o \Yeq_{Z/\gamma} \vevof{\sigma v}_{\pi_o  Z/\gamma \to \pi_+ \pi_-} = 
\Yeq_o\Yeq_\pm \vevof{\sigma v}_{\pi_o \pi_\pm \to Z/\gamma \pi_\pm} = 
{\Yeq_o}^2 \vevof{\sigma v}_{\pi_+ \pi_- \to \pi_o Z/\gamma}\,,
\eeq
and defined
\beq
\vevof{\sigma v}_{\pi_+ \pi_{-} \to \pi_o V} =
\vevof{\sigma v}_{\pi_+ \pi_{-} \to \pi_o \gamma}
+\vevof{\sigma v}_{\pi_+ \pi_{-} \to \pi_o Z}\,.
\eeq

For the $\su2$ case there is a single non-trivial 
chemical potential (\ref{eq:chem.pots}) and an associated conserved charge
\beq
q = Y_- - Y_+  \,.
\label{eq:def.of.q}
\eeq
Using $q$, the two independent Boltzmann equations become
\bea
\frac{dY_+}{dx} &=&  - \sqrt{\frac{\pi g(T)}{45 G}}\frac{M}{x^2}  \Biggl\{
\left[  Y_+ (Y_+ +q) -\Yeq_o{}^2 \right] 
\vevof{\sigma v}_{\pi_o\pi_o\to SM} 
+ \left[  Y_+ (Y_+ + q) -  Y_o^2 \right] \vevof{\sigma v}_{\pi_+\pi_{-}\to \pi_o\pi_o}
\cr && \qquad\qquad\qquad \qquad
+ \left[  Y_+ (Y_+ + q) -   Y_o \Yeq_o \right] 
 \vevof{\sigma v}_{\pi_+ \pi_{-} \to \pi_o V} ]  \Biggr\} \,,
\cr && \cr
\frac{dY_o}{dx} &=&
 - \sqrt{\frac{\pi g(T)}{45 G}}\frac{M}{x^2}  \Biggl\{
\left( {Y_o^2} - { \Yeq_o}^2\right) \vevof{\sigma v}_{\pi_o\pi_o\to SM} 
+\left [Y_o^2 -  Y_+ (Y_+  + q) \right] \vevof{\sigma v}_{\pi_o\pi_o\to \pi_+\pi_{-}}
\cr &&~
+ \left[  (2Y_+ + q) (Y_o - \Yeq_o)  - Y_+ (Y_+ + q) + Y_o \Yeq_o \right]
\vevof{\sigma v}_{\pi_+ \pi_{-} \to \pi_o V}    \Biggr\} \,.
\label{eq:dY2}
\eea
From (\ref{eq:def.of.ccal}) we find that
\bea
\Yeq_o \Yeq_{Z/\gamma} \vevof{\sigma v}_{\pi_o {Z/\gamma} \to \pi_+ \pi_{-}}&=&
\frac T{2 [2\pi^2 s(T)]^2} \int_{s_o}^\infty ds \, P K_V \sqrt{s} K_1(\sqrt{s}/T) \sigma^{Z/\gamma}\,,
\cr && \cr
\Yeq_o{}^2 \vevof{\sigma v}_{\pi_o \pi_o \to SM} &=& \frac T{2[2 \pi^2 s(T)]^2}\int_{s_o}^\infty 
ds \, \sqrt{s} P^2 K_1(\sqrt{s}/T) \sigma_{\pi_o \pi_o \to SM} \,,\cr &&
\eea
where $\sigma^{Z/\gamma}$ are given in (\ref{eq:sigmaV}) and $ P,\, K_V$ are defined in (\ref{eq:P.KV}).

\begin{figure}[thb]
\centering
\centerline{\includegraphics[height=6cm]{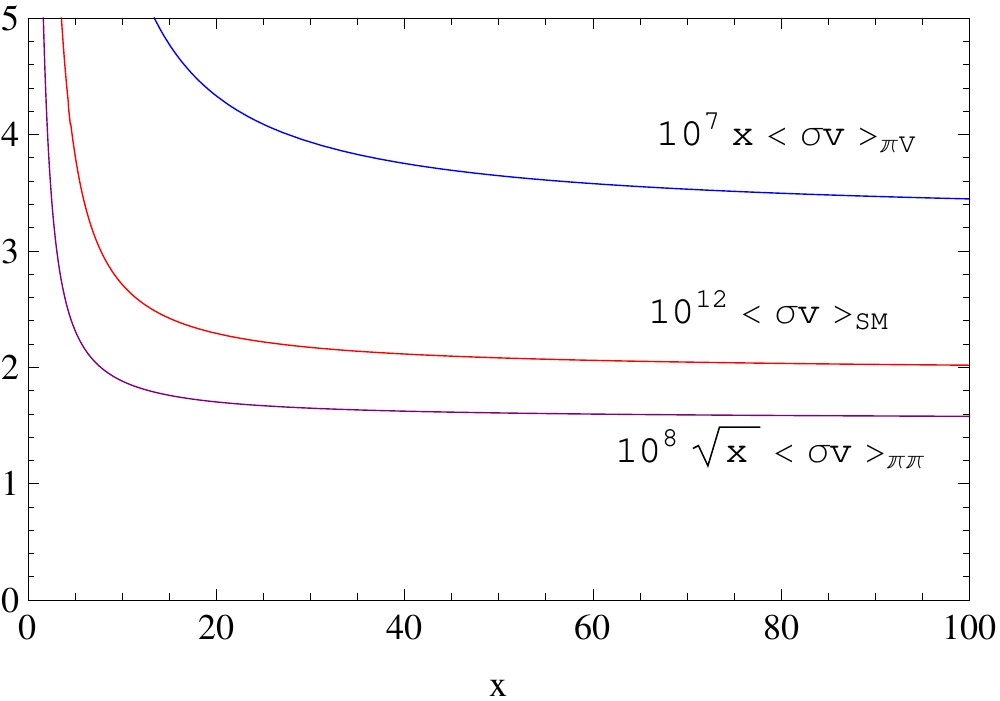}}
\caption{Cross sections for a representative set of parameters,
$(M,f,\lh, \lv)=(1000\,\gev,\, 950\,\gev, \,0.01,\,  0.63)$,
for which the model satisfies the cold-dark matter and direct-detection constraints.
Top curve: $10^7 x \vevof{\sigma v}_{\pi\pi\to \pi V}$;
middle curve: $10^{12} \vevof{\sigma v}_{\pi\pi\to SM}$;
bottom curve:  $ 10^8 \sqrt{x} \vevof{\sigma v}_{\pi\pi\to\pi\pi}$. 
The prefactors are chosen to fit the curves into the same graph 
and to illustrate the leading $x$ behavior.
All the cross sections are in $\gev^{-2}$.}
\label{fig:cs}
\end{figure}

\begin{figure}
$$
\includegraphics[height=5.5cm]{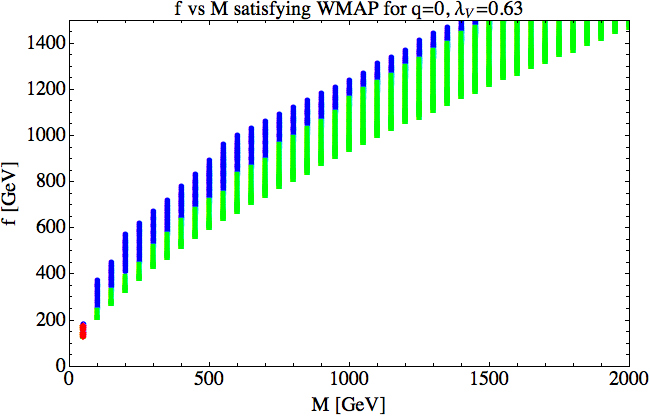}
$$
\caption{Region in the $M-f$ plane
allowed by the CMD constraint (\ref{eq:wmap.region})
when $ q =0 $, $ \lv=0.63 $, and $ |\lh|<1 $.
Blue points: subregion where  
$ \vevof{\sigma v}_{\pi\pi \to SM} (x=x_f) >
 \vevof{\sigma v}_{\pi\pi \to \pi V} (x=x_f) $.
Green points: subregion where
$ \vevof{\sigma v}_{\pi\pi \to SM} (x=x_f) <
 \vevof{\sigma v}_{\pi\pi \to \pi V} (x=x_f)$.
Red points: subregion excluded by the Higgs decay constraint (\ref{eq:higgs.constraint}).}
\label{fig:wmap.cs}
\end{figure}

The $\vevof{\sigma v} $ are plotted in Fig.\ref{fig:cs} for a representative parameter space point. 
The SM cross section is almost $x$-independent (corresponding to a predominance of $s$-wave scattering),
while the $ \gamma/Z $ cross section is proportional to $1/x$,
indicating a predominance of $p$-wave scattering. It is interesting to note that the
DMP$\to$DMP cross section has an unusual $1/\sqrt{x} $ behavior for large $x$
that results from all particles having the same mass and the amplitude being non-zero
and finite at threshold, which for this model is 
a consequence of the chiral couplings of the DMPs.  One can see,
that $\vevof{\sigma v}_{\pi\pi\to SM}$ is much smaller than 
$\vevof{\sigma v}_{\pi\pi\to \pi V}$ or $\vevof{\sigma v}_{\pi\pi\to\pi\pi}$
for the particular choice of parameters. The relevance of $\vevof{\sigma v}_{\pi\pi\to SM}$
can be understood by referring to Fig.\ref{fig:wmap.cs} where we
compare  $\vevof{\sigma v}_{\pi\pi\to SM}$ and
 $\vevof{\sigma v}_{\pi\pi\to \pi V}$ at the decoupling temperature (the point
at which the DMP particle density begins to deviate significantly from
its equilibrium value -- see Sec.\ref{sec:zero.q})
for points that satisfy the cold-dark matter (CDM)
relic-abundance constraint (see Eq.(\ref{eq:wmap.region}) below).

To obtain the particle densities and their freeze out temperatures it is necessary to solve a 
system of coupled linear differential equations for $\{ Y_o, Y_+ \}$ given by (\ref{eq:dY2}).
The boundary conditions are determined by requiring that at
low $x$ the DM sector is in equilibrium with the SM:
\beq
x< x_f: ~~
Y_o = \Yeq_o \,, \quad {\rm and}\quad
Y_\pm = \Yeq_\pm = \sqrt{\Yeq_o{}^2 + \frac{q^2}4} \mp \frac q2.
\label{eq:bc}
\eeq
Note that (\ref{eq:dY2}) and (\ref{eq:bc}) imply that both the equations and initial conditions are invariant under
$ Y_+ \leftrightarrow Y_- $ {\em and} $ q \leftrightarrow -q $.

For the following it is useful to note that $ \vevof{\sigma v}_{\pi_o \pi_o \to SM} $ 
depends on $ \lh $ only in the combination $ \lh/f^2 $, while 
$\vevof{\sigma v}_{\pi_+ \pi_- \to \pi_o V } $ depends on $ \lv $ only
as $ \lv/f^3 $. This implies that we can take
$M, f$ and $ \lh $ as independent parameters, fixing $\lv $ at
some convenient value as in (\ref{lambda'}); any other value of $ \lv$
can be obtained by
appropriate rescaling of $f$ and $ \lh $.

--
\subsection{Zero charge solutions}
\label{sec:zero.q}

When $q=0$ all DMP will have the same initial equilibrium distribution,
the relevant solutions to the BE then correspond to $ Y_{o,\pm} = Y $; substituting
this (and $q=0 $) in (\ref{eq:dY2}) we find 
\beq
\frac{dY}{dx} = - \sqrt{\frac{\pi g(T)}{45 G}}\frac{M}{x^2} \left( Y - \Yeq \right)  \Bigl\{
\left( Y + \Yeq\right) 
\vevof{\sigma v}_{\pi_o\pi_o\to SM}  
+ Y  \vevof{\sigma v}_{\pi_+ \pi_{-} \to \pi_o V} \Bigr\}\,,
\label{eq:dY.zero.q}
\eeq
where we drop the $o,\pm$ subindices.

Approximate solutions to this equation are readily obtained. We find that
to  good accuracy (see Fig.\ref{fig:cs})
the cross sections have an $s$ and $p$ wave behaviors 
for $ x> 10 $:
\beq
\vevof{\sigma v}_{\pi_o\pi_o\to SM}  \simeq \sigma_{SM}\,, \qquad
\vevof{\sigma v}_{\pi_+ \pi_{-} \to \pi_o V}  \simeq \inv x \sigma_V \,,
\label{eq:sv.approx}
\eeq
where $ \sigma_{SM,V} $ are approximately $x$-independent.

Near the decoupling temperature we write $ Y = \Yeq+ \Delta $
and neglect terms proportional to $ d\Delta/dx$ and $ \Delta^2 $; 
then (\ref{eq:dY.zero.q}) becomes
\beq
\Delta \simeq\frac{x^2}{2\vt + \vtt /x}
\,; 
\quad \vt = \sqrt{\frac{\pi g(T)}{45 G}}{M\sigma_{SM}} \,,
\quad \vtt = \sqrt{\frac{\pi g(T)}{45 G}}{M\sigma_{V}} \,,
\label{eq:cs.app}
\eeq
where we also approximated $ d\Yeq/dx \simeq - \Yeq $.

For large $x$, in contrast, $ \Delta \simeq Y \gg \Yeq $ and (\ref{eq:dY.zero.q})
becomes
\beq
\frac{d\Delta}{dx} = -  \frac{\vt x +  \vtt}{x^3}   \Delta^2
\then \Delta_\infty  \simeq \frac{x_f^2}{\vt x_f + \vtt/2} \,.
\eeq
where $ 1/\Delta(x_f)$ is neglected.

Finally the decoupling `temperature' $x_f $ is obtained from the condition $ \Delta(x_f) = 
c \Yeq(x_f) $, where $c$ is a numerical constant. This gives
\bea
Y_\infty &=&\frac{x_f^2}{\vt x_f + \vtt/2} \,, \cr
x_f &=& \ln \left[ ac(c+2)\vt\xi^{-1/2} + a c (c+1) \vtt\xi^{-3/2}
\right]\,; \quad \xi = \ln[c(\vt + \vtt) a]
\label{eq:Y&xf}
\eea
where $a$ is defined in (\ref{eq:Yeq.def}) and $\vt,\,\vtt$ in (\ref{eq:cs.app});
this result is better suited for the case $ \vtt \gg \vt $ than the one
presented in \cite{Kolb:1990vq}. We will follow this reference
 and choose $ c(c+2) = 1 $ or, $ c \simeq 0.414 $.
In calculating the relic abundance it is important to remember that
$ Y_\infty$ refers to {\em each} DMP species, so that 
the total abundance will be proportional to $3 Y_\infty $.

\begin{figure}[tbh]
$$\includegraphics[height=6cm]{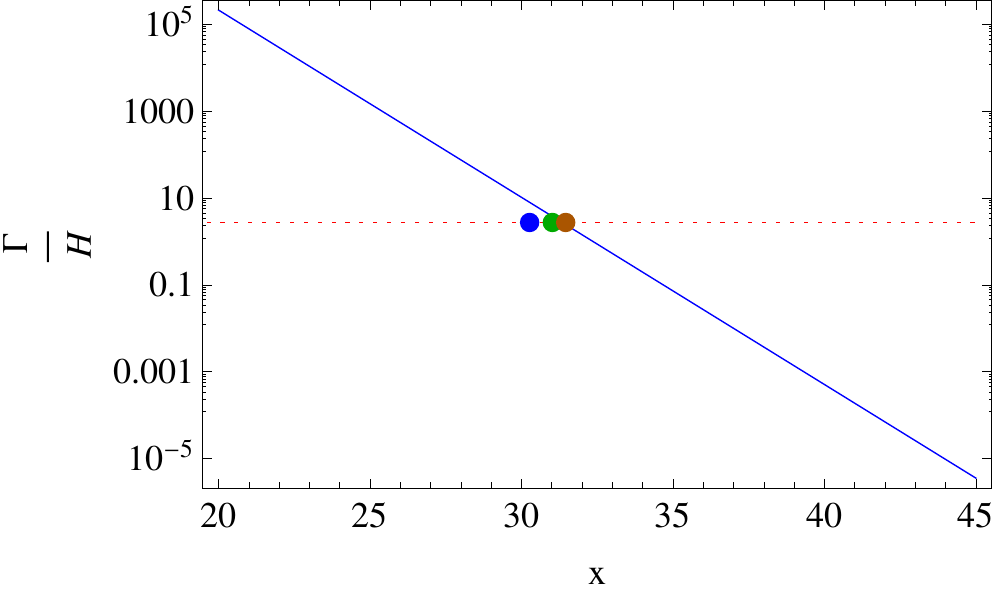} $$
\caption{Plot of $ \Gamma/H $
for the same parameters as in Fig.\ref{fig:cs}.
 We also include the values of $ x_f $ obtained from the condition $  \Delta = c \Yeq$ for 
$c=0.414, \, 0.732,\, 1$ (left, center and right heavy dots on the dashed line, respectively). 
The freeze-out condition $ \Gamma = H $ corresponds to $ x_f \simeq 31.3$
which coincides almost exactly with the $c=1$ value.}
\label{fig:gamma.h}
\end{figure}

An alternative definition of $x_f$ can be derived by assuming
$ Y $ is close to $ \Yeq $ and casting
(\ref{eq:dY.zero.q}) in the form
\beq
\frac x{\Yeq_o} \frac{dY}{dx} = - \frac\Gamma H \left( \frac Y{\Yeq_o}-1 \right) \,; \qquad
\frac\Gamma H = \left( \frac{ 2 \vt + \vtt}x  \right)\Yeq_o 
\eeq
so $x_f$ can be defined as the point where $ \Gamma/H = 1 $. A plot of $ \Gamma/H $
for representative values of the parameters, and a comparison with
the previous definition of $ x_f $ is given in Fig.\ref{fig:gamma.h}. 
This also illustrates that $x_f$ in general is large enough
for the approximations (\ref{eq:sv.approx}) to be valid.

\begin{figure}[tbh]
\centering
\centerline{\includegraphics[height=6.5cm]{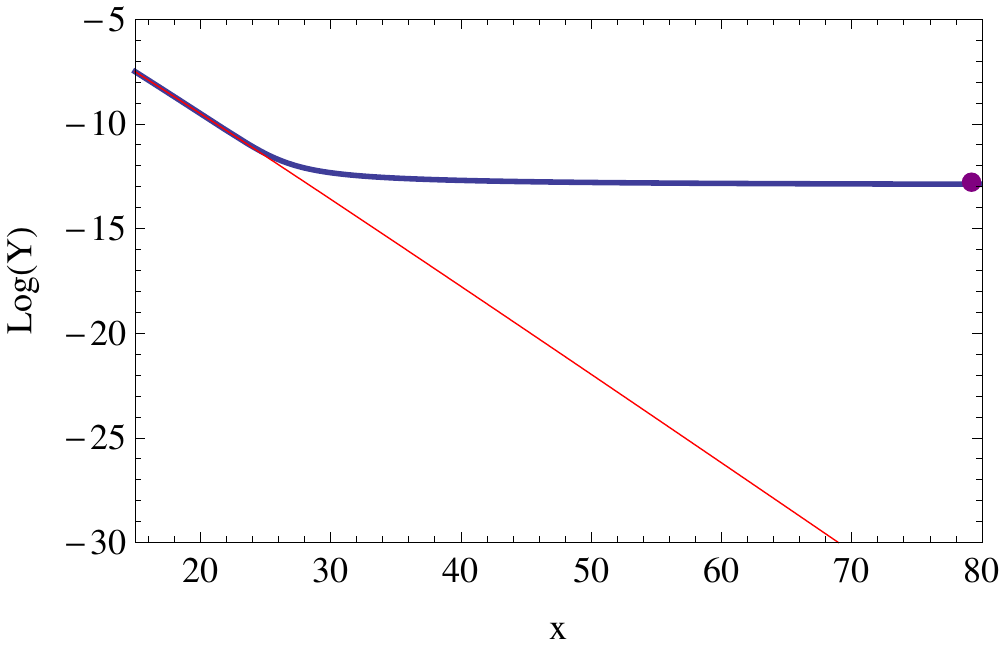}}
\caption{Plot of the yield $Y$ as a function $x$ 
for the representative point of Fig.\ref{fig:cs} when $q=0$.
Dark matter pion aboundance is depicted in blue, and the equilibrium distribution is shown in red. The heavy dot on the right indicates the value of $ Y_\infty$ obtained form (\ref{eq:Y&xf}) using $c=1$. All masses are in GeV.} 
\label{fig:beq0}
\end{figure}

In Fig.\ref{fig:beq0} we
compare the relic abundance derived numerically 
with the one obtained from (\ref{eq:Y&xf}), showing that, at least in
this instance, the latter is reasonably accurate.
From this figure one can also see that  the decoupling point inferred from 
the numerical solutions equals the analytically obtained values
within 10\%.

\subsection{Behavior for small values of $|q|$ \label{sec:non.zero.q}}

We now turn to the case where $q $ is small but non-vanishing. In this case  it is convenient to define
\bea
Y_t &=& Y_o + Y_+ + Y_- = Y_o + 2 Y_+ + q \,,\cr
&& \nonumber \\
Y_d &=& \frac{Y_+ + Y_-}2 - Y_o = Y_+ - Y_o + \frac q2 \,,
\eea
in terms of which Eqs.(\ref{eq:dY2}) become
\bea
Y_t' &=&  -\inv3 ( y_t^2 + 2 y_d^2 )( A+B) +( y_t +  y_d^2) B + \Biggl[ \frac{q^2}4 (2A+B) +3A \Biggr] \,,
\cr
Y_d' &=&  \inv3 y_d(y_d-2 y_t) \left( A +B+ \frac32 C \right)
- y_d( y_d  + 2 ) B + \frac{q^2}4 \left( A +3B+ \frac32 C\right) \,,
\label{eq:dY3}
\eea
where $y_{t,d} = Y_{t,d} /\Yeq_o $ and
\beq
\{ A,\, B,\, C\}  = \sqrt{\frac{\pi g(T)}{45 G}}\frac{M}{x^2} \Yeq_o{}^2
\{ \vevof{\sigma v}_{\pi_o\pi_o\to SM},\; \vevof{\sigma v}_{\pi_+ \pi_{-} \to \pi_o V} 
,\;\vevof{\sigma v}_{ \pi_o\pi_o\to\pi_+\pi_{-}} \}\,,
\eeq
while the initial conditions (\ref{eq:bc}) correspond to
\bea
Y_t  = \Yeq_t &=&  \Yeq_o + 2 \sqrt{{\Yeq_o}^2 + q^2/4} \,,\cr
Y_d  = \Yeq_d &=&  -\Yeq_o +  \sqrt{{\Yeq_o}^2 + q^2/4} \,.
\eea

Now $Y_{t,d}$ are even in $q$, and assuming they are analytical 
in $q$ it follows that they depend on $q^2 $; at $q=0 $, we have $Y_t = 3Y$ and $Y_d=0$.
Taking a derivative of (\ref{eq:dY3}) with respect to $q^2$ and evaluating at $ q =0 $ gives
\bea
\dqq{Y_t}' &=&  -\frac{2y}{\Yeq_o} \left( A+B - \frac B{2y} \right) \dqq{Y_t} +  \frac{2A+B}4 \,,
\cr
\dqq{Y_d}' &=&  -\frac{2y}{\Yeq_o} \left( A +B + \frac By+ \frac32 C \right) \dqq{Y_d}+ \frac{A +3B+ 3C/2}4 \,, 
\eea
where $y=Y_o/\Yeq_o $. 
Initially, 
\bea
&& \dqq{Y_t} = \dqq{\Yeq_t} = \inv{4 \Yeq_o} \,,\cr
&& \dqq{Y_d} = \dqq{\Yeq_d} = \inv{8 \Yeq_o} \,.
\eea

Now, a differential equation of the form
\beq
Z' = u Z + v 
\label{eq:dif-eq}
\eeq
has solution
\beq
Z(x) = \int_{x_i}^x ds \, v(s) \exp \left[ \int_s^x dr \, u(r) \right] + 
Z_i  \exp \left[ \int_{x_i}^x dr \, u(r) \right]\,, \qquad
Z_i = (x_i) \,.
\label{eq:dif-eq-sol}
\eeq
In particular, if $ v(x) > 0 $ for all $x$, and  $ Z_i > 0 $, then $Z(x)>0$ for $ x > x_i $.
Applying this to $Z=(\partial Y_{t,d}/\partial q^2)_{q=0}$, that have initial values
$ \sim1/\Yeq_o(x_i) >0 $, we find that
\beq
\dqq{Y_{t,d}} > 0 \,, ~ {\rm for}~ x \ge x_i \,.
\eeq

The relic abundance is obtained from the expression~\cite{Kolb:1990vq}
\beq
\Omega_{\rm DM} h^2 = 2.7711 \times 10^8 (M/\gev)(Y_o + Y_+ + Y_-)_{x=\infty}
= 2.7711 \times 10^8 (M/\gev)\, Y_t|_{x=\infty} \,,
\label{omega}
\eeq
since $ Y_t (q\not=0) > Y_t(q=0) $ (at least for small $q$
and with the other parameters fixed), it follows that
\beq
\Omega_{\rm DM} ( f, M, \lh, \lv; q =0 ) < \Omega_{\rm DM} ( f, M, \lh, \lv;q \not= 0 )\,.
\eeq
If $ \Omega_{DM} ( f, M, \lh,\lv; q =0 ) < \Omega_{\rm CDM} $
for some parameters $ \{ f,\, M ,\, \lh,\, \lv \}$, then there will be a non-zero $q$ such that
$ \Omega_{\rm DM} ( f, M, \lh, \lv;q  ) = \Omega_{ \rm CDM} $. 
That is, if the predicted abundance falls below the observations when $q=0$, one can
always ``make-up'' the difference by introducing an appropriate $q$ (at least
when the difference is small). It follows that the
the region in parameter space that can satisfy the CDM constraints is determined by
\beq
\Omega_{DM} ( f, M, \lh, \lv;q =0 ) < \Omega_{\rm CDM} \,.
\label{eq:as.dm}
\eeq
A non-zero value of $q$ does not, of course, affect the direct-detection
probability.

\begin{figure}[th]
\centering
\centerline{\includegraphics[height=6cm]{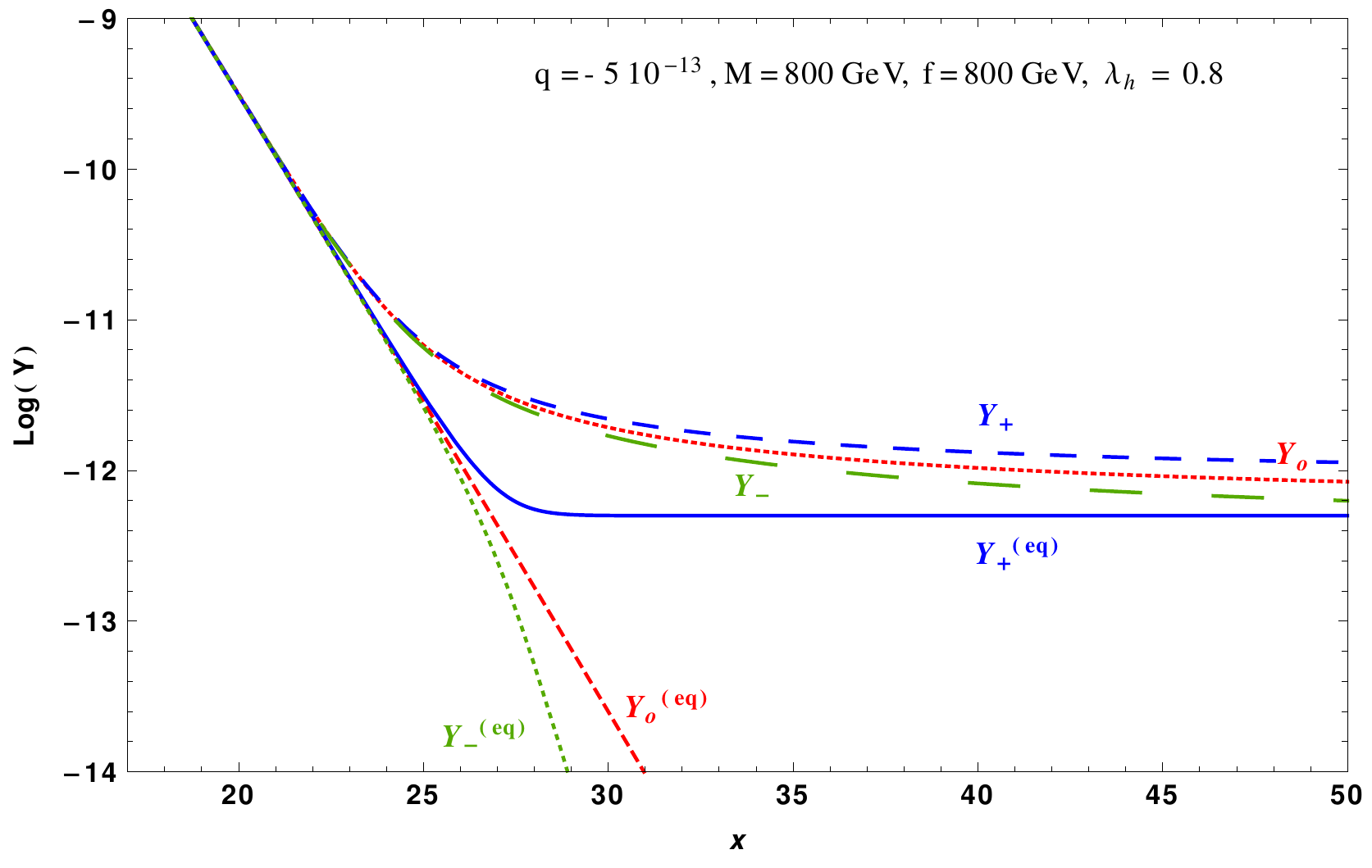}}
\caption{Illustration of the $q \neq 0 $ case.}
\label{BEQ-q}
\end{figure}

We illustrate Boltzmann equation solutions for small $q$ in Fig.\ref{BEQ-q}. In general, there is a small range of $|q| \sim 10^{-12}-10^{-13}$ for which 
differences among the $Y_+$, $Y_-$ and $Y_0$ abundances and between these and their
equilibrium values are easily distinguished (it these cases
the freeze-out temperatures for all three DMP components are very close). For smaller values, the effect of $q$ is negligible,
while for larger values the effects of $q$ dominate the relic abundance and we find that $ Y_o + Y_+ + Y_- \simeq |q|$.

\section{Experimental limits on model parameters \label{sec:exp.const}}

\subsection{Constraints from the cold dark matter (CDM) relic density measurements}
\label{sec:CDM.constraints}

In this section we will obtain the numerical solution to the
Boltzmann equations
for the case $ q =0 $, when\footnote{Note that for $q=0$ case,
DMP $\to$ DMP scattering cross sections do not enter
Eq.(\ref{eq:dY.zero.q}).} $Y_+ = Y_- = Y_o = Y$,
 and find the region of parameter space that meets
the relic-abundance constraint \cite{pdg}~\footnote{The
range we use corresponds to the WMAP results; the PLANCK 
constraints $0.112 \leq \Omega_{\rm DM} h^2 \leq 0.128$ \cite{PLANCK},
though more stringent, do not lead to significant
changes in the allowed regions of parameter space.}
\beq
0.094 \leq \Omega_{\rm DM} h^2 \leq 0.130 \,.
\label{eq:wmap.region}
\eeq

As noted at the end of Sec.\ref{sec:su2.BE} the solutions will depend on
3 independent parameters that we choose as $M$, $f$ and $ \lh$; 
without loss of generality, we fix $ \lv$ to the value (\ref{lambda'}).
We scan the 3-dimensional parameter space $(M,f,\lh)$ in the
ranges $ 50\,\gev\le M \le 2\,\tev$, $ 50\,\gev\le f \le 1.5\,\tev$, $ 10^{-4} \le |\lh|
\le 1 $ for points allowed by (\ref{eq:wmap.region});
we also impose the constraint (\ref{eq:f-M}) and the one derived from 
$ h\to\pi\pi $ decay, which is open in the low $M$ region (cf. Sec.\ref{sec:hpp});
note that in this
region of parameter space the decay $ Z \to\pi\pi\pi $ is kinematically forbidden, so that the restriction (\ref{eq:z3pi}) does not apply. The 
$ q \not=0 $ case is included by considering only the
upper inequalities (see (\ref{eq:as.dm})).
In the next section we consider the constraints direct-detection results
from XENON100 and XENON1T experiments \cite{Xenon}. In particular,
using $ Y_o + Y_+ + Y_- \simeq |q|$ for $ q \gg 10^{-12} $
(cf. the end of Sec. \ref{sec:non.zero.q}) we fin that
(\ref{eq:as.dm}) satisfies (\ref{eq:wmap.region}) provided
\beq
\frac{3.4 \times 10^{-10}}{M/\gev} < |q| <  \frac{4.7 \times 10^{-10}}{M/\gev}
\qquad M \ll 100 \gev
\label{eq:cdm.q}
\eeq

\begin{figure}[th]
$$
\includegraphics[height=5.5cm]{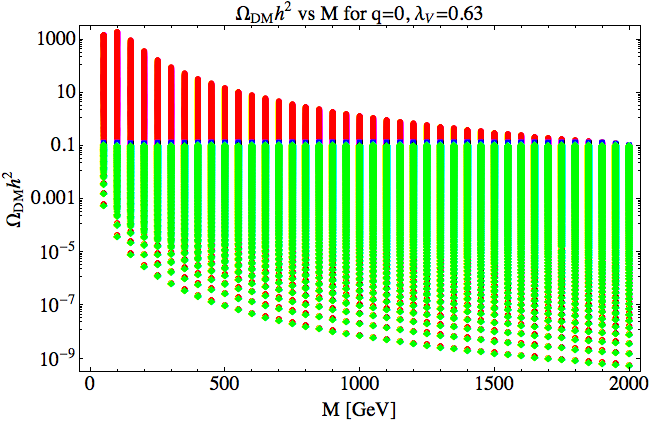}
\qquad
\includegraphics[height=5.5cm]{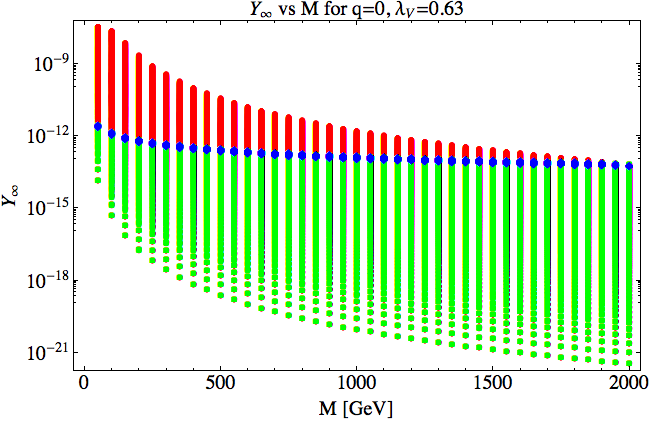}$$
$$
\includegraphics[height=5.5cm]{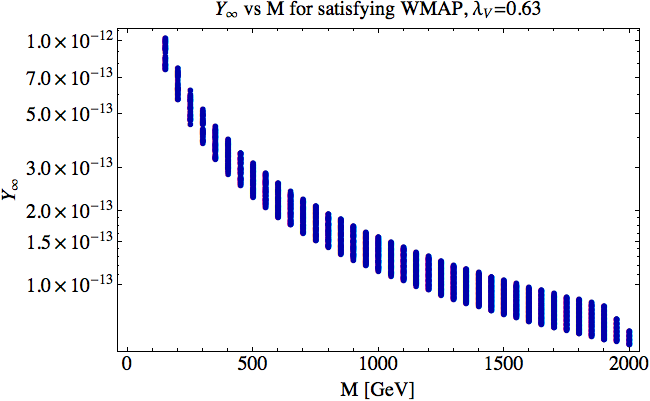}
$$
\caption{$\Omega_{\rm DM}h^2$ (top left) and $Y_\infty$ (top right)
dependence on the DMP mass $M$ for all
values of $f,\, \lh$ in the region scanned, and when $ q=0 $
and $ \lv = 0.63$.
Red points: DM over-abundance ($ \Omega_{\rm DM}h^2 > 0.13$); 
blue points: region allowed by the CDM constraint (\ref{eq:wmap.region});
green points: DM under-abundance ($ \Omega_{\rm DM}h^2 < 0.094 $), which are allowed for appropriately chosen non zero $q$. The CDM-allowed region
for $ Y_\infty$ is amplified in the bottom panel in order to better
see the dependence on $M$. } 
\label{OmegaM}
\end{figure}

\begin{figure}[thb]
$$
\includegraphics[height=5.5cm]{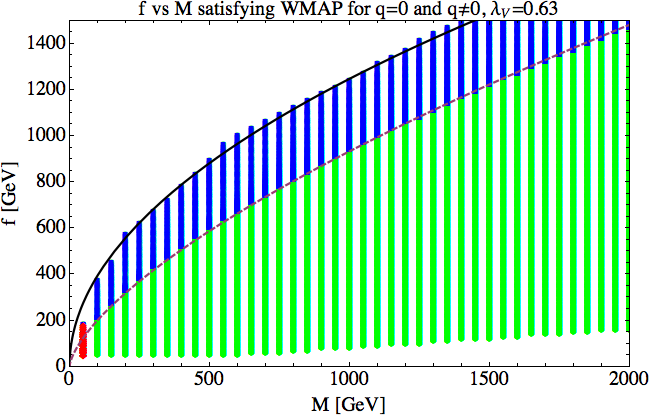}
\includegraphics[height=5.5cm]{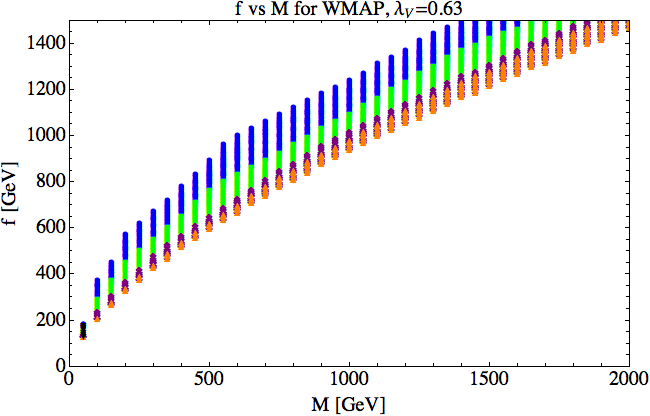}
$$
\caption{Left panel: region in the $f-M$ plane allowed by the
CDM constraint
(blue); the region corresponding to DM under-abundance (green);
and the region excluded by the Higgs decay constraint
Eq.(\ref{eq:higgs.constraint}) (red).  The
solid and dashed  black line correspond  to the 
analytic approximations (\ref{eq:analyt}). Right panel: $ \lh$
dependence of the points in the region allowed by (\ref{eq:wmap.region}).
Blue: $0.0001 \le \lh \le 0.01$, green: $0.01 \le
\lh \le 0.3$, purple: $0.3 \le \lh \le 0.6$, orange:
$0.6\le \lh \le 1$.
Red points are disallowed by (\ref{eq:higgs.constraint}).}
\label{M_f}
\end{figure}

In Fig.\ref{OmegaM}  we plot the relic abundance 
 $\Omega_{\rm DM}h^2$ and low-temperature distribution $Y_{\infty}$ as functions
of $M$. In Fig.\ref{M_f} we show the region in the $M-f$ plane allowed by the
CDM constraint (\ref{eq:wmap.region}) as well as the region allowed by 
$ q \not=0 $. Note, from the bottom panel of this figure, that
$ Y_\infty $  cannot be assumed to be $M$ independent as usually assumed in many models.

In Fig.\ref{M_f} we present the region in the $M-f$ plane
allowed by the CDM constraint 
We see from that figure that $ \Omega_{\rm DM}$ increases with $ \lh $:
and the region of sufficiently small (large) $ \lh $ corresponds
to an under (over)-abundance of DM. This is in contrast to models
where the leading coupling to the DM fields is through the Higgs-portal
interaction~\cite{higgs-portal}. We trace this difference to the presence of the 
$ \pi\pi\to Z \pi $ interaction: comparing Fig.\ref{fig:wmap.cs} and  Fig.\ref{M_f} we see that the region where the  relic abundance 
is small (but still allowed by the data) corresponds to small values of $\lh $
and also to $ \vevof{\sigma v}_{\pi\pi \to SM} (x=x_f) >
 \vevof{\sigma v}_{\pi\pi \to \pi V} (x=x_f)$; while large values of
$ \lambda _x $ correspond to the larger allowed values of the relic abundance
and to $ \vevof{\sigma v}_{\pi\pi \to SM} (x=x_f) <
 \vevof{\sigma v}_{\pi\pi \to \pi V} (x=x_f)$. 

The $q=0 $ allowed region in Fig.\ref{M_f} can be approximated analytically
by
\beq
39.65 \,\sqrt{  M }\ge f \ge 9.33\, M^{2/3} \qquad
(M,f\; {\rm in}\; \gev;\;   M<2\,\tev,\, |\lh| \le 1,\, \lv=0.63) \,.
\label{eq:analyt}
\eeq
We now use this result to extend the CDM limits  with reasonable accuracy to the 
whole region of parameter space of interest.
To do that note first that the $s$-wave contribution to  $ \vevof{\sigma v}_{\pi\pi \to SM} $
is generated by the $ \pi\pi \to hh $ contribution (cf. Eq.(\ref{eq:dmp-sm}))
so that in (\ref{eq:sv.approx}) $ \sigma_{SM}
\sim ( \lh M/f^2)^2  $  where the factor $ (|\lh|/f^2)^2 $
comes from the vertices, while the factor of $M^2$ is needed 
to get the right units (the other mass scales can be ignored for $ M > m_h/2 $). 
Similarly $ \sigma_V \sim ( \lv M^2/f^3)^2  $  where
the factor $ (|\lv|/f^3)^2 $
comes from the vertices, while the factor of $M^4$ is needed 
to get the right units. 

Using this in (\ref{eq:Y&xf}) and (\ref{omega}) we find that up to a weak
logarithmic dependence the parameters, $1/(h^2 \Omega_{DM})$ will 
depend on a linear combination of $ (\lh M/f^2)^2 $ and $ (\lv M^2/f^3)^2 $. 
Comparing then Fig.\ref{fig:wmap.cs} and Fig.\ref{M_f} we find that the
upper limit in (\ref{eq:analyt}) corresponds to parameters where $ \sigma_{SM} $
dominates  and where the upper limit in (\ref{eq:wmap.region}) is saturated; while
the lower limit in (\ref{eq:analyt}) corresponds to parameters where $ \sigma_V $
dominates  and where the lower limit in (\ref{eq:wmap.region}) is saturated.
Using this in conjunction with (\ref{eq:analyt}) we find that the CDM constrain
reduces to
\beq
4.04 \times 10^{-7} \le \left( \frac{\lh \, M}{f^2} \right)^2
+ 0.93  \left( \frac{\lv \, M^2}{f^3} \right)^2 \le 5.59 \times 10^{-7} \delta_{q0}
\quad (M,f\; {\rm in}\; \gev)\,.
\label{eq:CDM.constraint}
\eeq
where $\delta_{q0}$ vanishes when $ q \not= 0 $ 
so that there is no upper limit in (\ref{eq:CDM.constraint}) in this case.

\subsection{Direct detection constraints}

\begin{figure}[thb]
\centering
\centerline{\includegraphics[height=4cm]{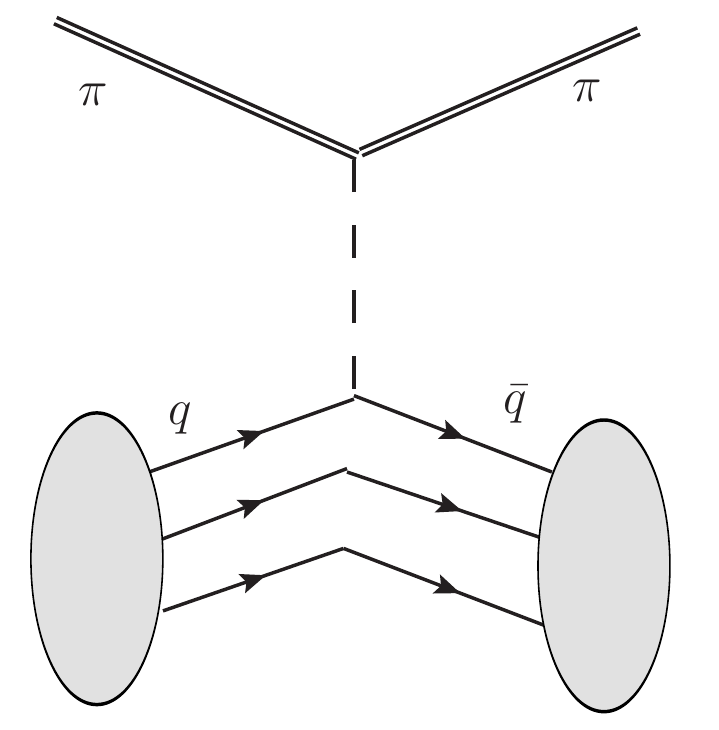}}
\caption{Direct detection process.}
\label{DD_process}
\end{figure}

The direct detection experiments probe the elastic scattering of DM particles off
different kinds of materials~\cite{DAMA,CDMS,Xenon}. For the present model the leading interaction is the
$ \pi N \to \pi N $ scattering of DMP off the material's nucleons $ N $
(Fig.\ref{DD_process}) through a $t$-channel Higgs exchange. The corresponding
hard process was discussed in Sec.\ref{sec:direct.detection} where we show
that the DMP-quark scattering cross section  (\ref{eq:direct.detection})
is proportional to $ (\lh M^2/f^2)^2 $
 
\begin{figure}[thb]
$$ \includegraphics[height=7cm]{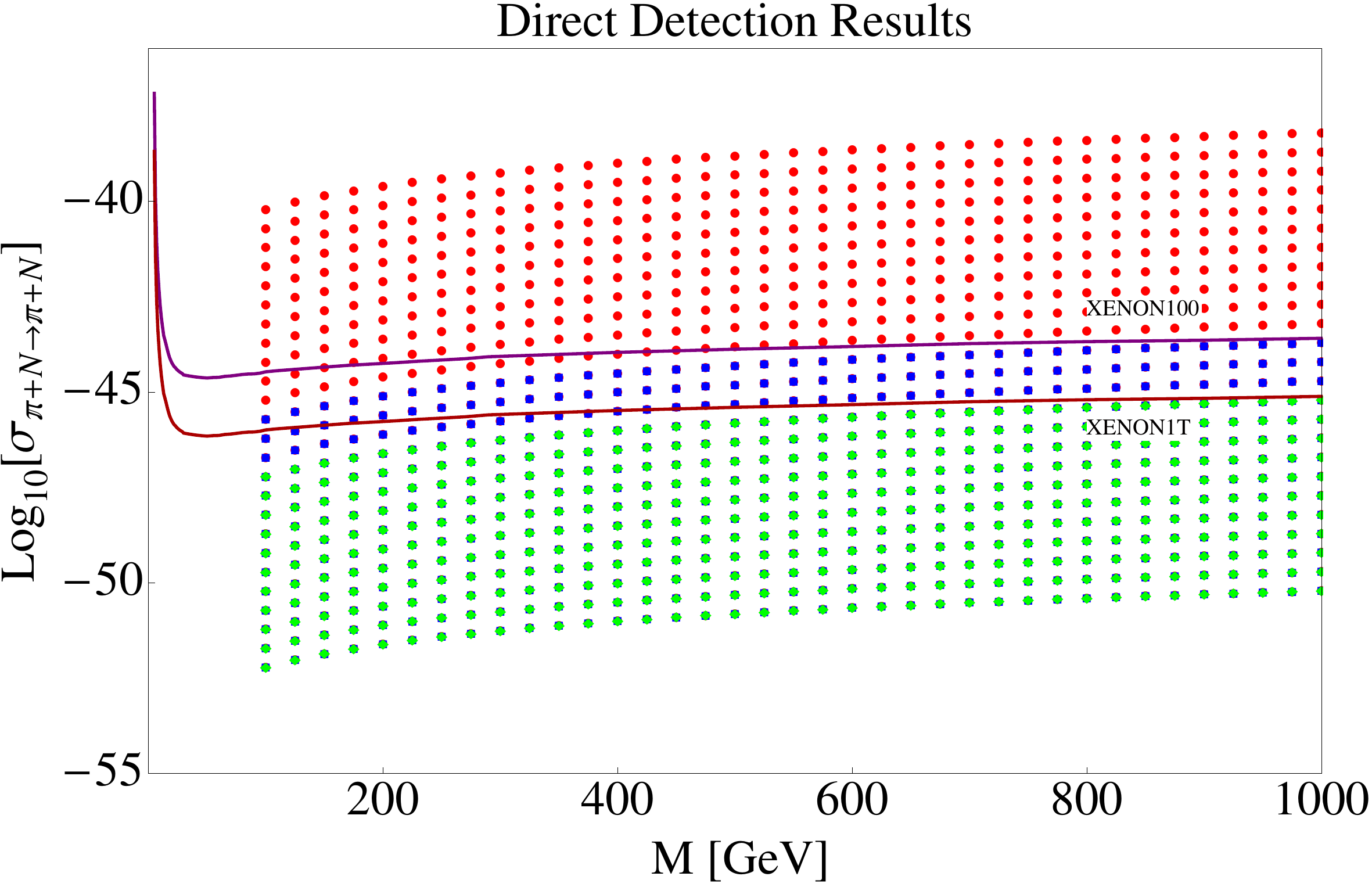}$$
\caption{Direct detection constraints from XENON experiments. 
XENON100 excludes all points above the solid line 
in purple at the top, which corresponds to the constraint
$ \lh/f^2 < 10^{-5.5} $. XENON1T is projected
to exclude all points above the lower (red) solid line
and would correspond to the constraint 
$ \lh/f^2 < 10^{-6.5} $. }
\label{DD_bound}
\end{figure}

The parton-level interaction is converted to the nucleon level by using effective nucleon ${f_q}^N$ $(N = p,n)$ couplings defined as~\cite{Belanger:2008sj}
\begin{eqnarray}
\langle N|m_q\bar{\psi_q}\psi_q|N \rangle={f_q}^N M_N \,,
\end{eqnarray}
where $M_N$ is the nucleon mass and
$f_u^p= 0.0160,\; f_d^p= 0.0193,\; f_s^p= 0.0410$, for the proton;  $f_u^n= 0.0108,\; f_d^n= 0.0284,\;f_s^n= 0.0409$
for the neutron; while for the heavy quarks the $f_q^N$ are generated by gluon exchange with the nucleon and are given by
 \begin{eqnarray}
f_Q^N=\frac{2}{27} \left (1-\sum_{q=u,d,s} f_q^N  \right ) \, \quad Q=c,t,b.
\label{eq:fQ}
\end{eqnarray}
Then, DMP scattering with a nucleon composed of $Z$ protons and $A-Z$ neutrons is \cite{Belanger:2008sj}
\begin{eqnarray}
\sigma_{\pi N}= \frac{1}{\pi}\left ( \frac{m_N}{m_N + M} \right )^2 ( Z f^p + (A-Z) f^n)^2 \,; \qquad
\frac{f^N}{m_N} = \sum_q \frac{f_q^N}{m_q} \alpha_q
\label{ddcross}
\end{eqnarray}
and the sum is over all quarks.
The
$\alpha_q$ are effective couplings of DMP with the $q$-quarks, ${\cal L} = -\half \alpha_q \overline{\psi_q} \psi q \pi \pi$
that can be read off (\ref{eq:leff.psi.pi}):
\begin{eqnarray}
\alpha_q = \sqrt{2} \frac{m_q\,M^2}{m_h^2}  \frac{\lh}{f^2} \,.
\end{eqnarray} 

\begin{figure}[thb]
$$
\includegraphics[height=5.5cm]{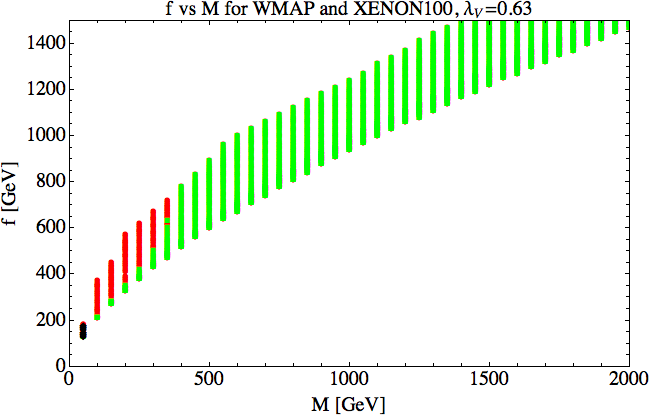} \quad
\includegraphics[height=5.5cm]{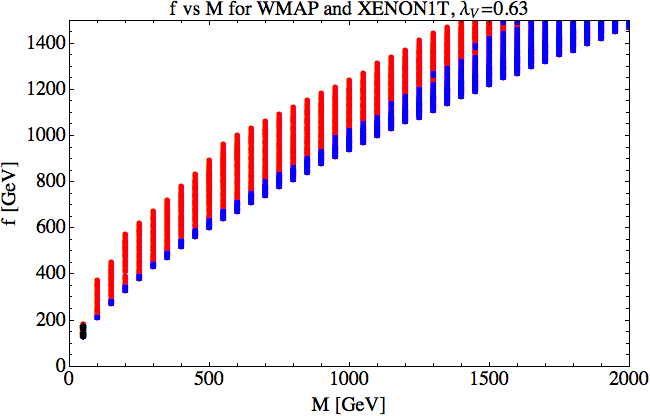}
$$

\caption{Left: region in the $M-f$ plane allowed by the CDM constraint
and allowed (green) or disallowed (red) by the XENON100 data (\ref{eq:xe.constraint}); 
black points are 
disallowed by (\ref{eq:higgs.constraint}). Right: same for the
predicted XENON1T exclusion region in red and allowed in blue. We took $ q =0$, $\lv=0.63$
and $|\lh|<1 $. }
\label{fig:xe.and.cdm}
\end{figure}

Using {\tt microOMEGAs} \cite{Belanger:2008sj} we evaluate numerically the DMP-nucleon scattering cross section for direct detection and then compare these results to the XENON100 and XENON1T bounds. The results are presented in Fig.\ref{DD_bound}. As indicated above, if $M$ is fixed the cross section depends only on $ \lh/f^2 $ and, in fact, the XENON bounds give rather simple expression for the constraints on this ratio:
\bea
{\rm XENON100}:&& f^2/\lh > 10^{5.5} \,,\cr
{\rm XENON1T}:&&f^2/\lh > 10^{6.5} \,.
\label{eq:xe.constraint}
\eea
The corresponding restrictions on the $M-f$ plane
over the CDM constrain are presented in  Fig.\ref{fig:xe.and.cdm}.

\subsection{Combined constraints on DMP model}

The parameters in the model are constrained by the relations
(\ref{eq:pert.constraint}), (\ref{eq:higgs.constraint}), (\ref{eq:CDM.constraint}), 
and (\ref{eq:xe.constraint}) that we collect here for convenience:
\bea
{\rm perturbativity:}&& f \ge  {\rm max}\{ \sqrt{4 \pi \lv}\,,\, 1 \}  \frac M{4\pi} \cr
{\rm Higgs~decay:}&& f > 5.9  |\lh|^{1/2} |7812.5 - M^2|^{1/2}
\left[ 1 - \left(\frac M{62.5} \right)^2 \right]^{1/8} \qquad (M < 62.5 \; [{\rm GeV}]) \cr
{\rm XENON100:}&& f > 562.3 |\lh |^{1/2} \cr  && \cr
{\rm CDM:}&& 4.04 \times 10^{-7} \le \left( \frac{\lh \, M}{f^2} \right)^2
+ 0.93  \left( \frac{\lv \, M^2}{f^3} \right)^2  \le 5.59 \times 10^{-7} \delta_{q,0}  \;,
\label{eq:all.constraints}
\eea
where $f,M$ are in \gev, and we used the XENON100 limit. The $ \delta_{q,0} $
factor indicates that the corresponding limit disappears when non-zero values of $q$ 
are allowed. 

\begin{figure}[thb]
$$
\includegraphics[height=6cm]{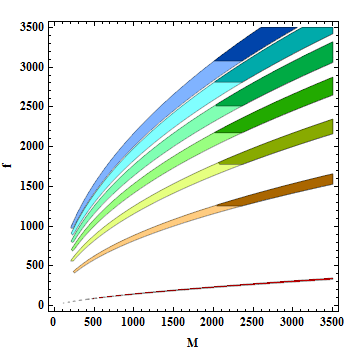} \quad
\includegraphics[height=6cm]{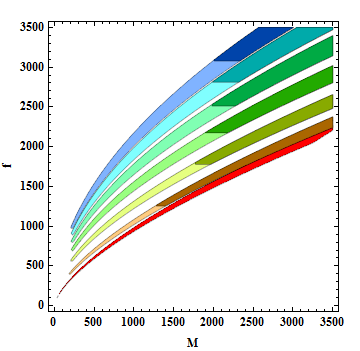}
$$
$$
\includegraphics[height=6cm]{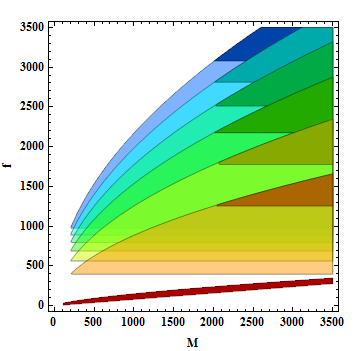} \quad
\includegraphics[height=6cm]{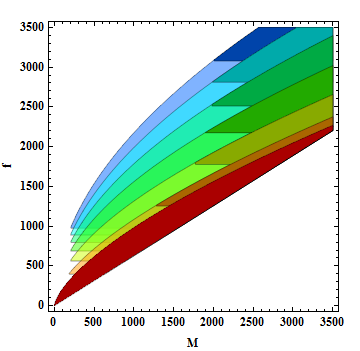}
$$
\caption{Top left panel: region in the $f-M$ plane allowed by the
combined constraints (\ref{eq:all.constraints}) when $ q=0 $ for $ \lv = 0.0023 $.
The various bands correspond 
to $\lh=\{ 0,\, 0.5,\, 1,\, 1.5,\, 2,\, 2.5,\,3 \}$ from bottom to top, respectively;
the darker regions correspond to those allowed by XENON1T.
Top right panel: same for $ \lv = 0.63$.
Bottom panels: same as the top panels when $ q \not=0 $.}
\label{fig:all.constraints}
\end{figure}

The resulting allowed regions in parameter space are
given in Fig.\ref{fig:all.constraints} for our benchmark value of $ \lv =0.63 $
as well as for the smaller natural value $\lv = 0.0023 $ derived by NDA (see Sec.\ref{sec:parameters}). 
As can be seen from this figure if $ \lh \not\simeq 0 $ 
current data excludes DMP masses below $ \sim 100 \,\gev $ while XENON1T would
push this limit above $ 1 \,\tev $. These limits do not apply when $ \lh \simeq 0 $;
in this case low values ($ < 100 \,\gev $) for $M$ and $f$ are allowed;
in this case a non-zero value of $q$ can always be found that meets all constraints
(see Eq. \ref{eq:cdm.q}).


\section{\label{sec-6}Conclusions \label{sec:concl}}

We have studied a phenomenological model, where dark matter particles are pseudo-Goldstone 
bosons associated with the spontaneous breaking $\gdm \to \hdm $; we refer to these
particles as dark matter ``pions''. The self-couplings and the couplings to the SM 
for such pionic DM differ from those
of conventional scalars due to their chiral nature.
We  have illustrated the formalism for the case $ \gdm = \su2\times\su2 $, $\hdm=\su2 $
for which we have calculated all possible interactions and solved the Boltzmann equations 
to study the thermal history of such pionic dark matter. We have also derived 
approximate analytic solutions and shown that they are consistent with the numerical calculations.

Our model of pionic dark matter satisfy relic abundance and direct detection constraint in a large region
of parameter space. When the coupling to the Higgs is not too small the 
DMP mass $M$ is required to lie above $ \sim 100 \,\gev $, and this lower limit will
increase to $ \sim 2 \,\tev $ if XENON1T does not detect a signal, since 
the absence of direct detection corresponds to relatively large values of $ f^2/\lh $.
For each value of $M$ the DMP decay constant $f$ is moderately constrained to a
range of values which is $ \sim 200 \,\gev $ wide.

Collider signature of such dark matters at LHC is hard to see. The channel  to study 
is essentially jets with missing energy~\cite{lhc}, which is similar to many other dark matter model signatures~\cite{dm.at.lhc}.
This requires a careful analysis to see if the existing bound in such channels put further constraints on the DMP 
parameter space, which lies beyond the scope of this paper. We will consider this in a future publication.

The DM couples to the SM via $Z$, $\gamma$ and $h$, therefore it does not
distinguish between fermion flavors. In particular there is no
mechanism for suppressing the effects of the $\pi$
at XENON experiments and enhancing them at DAMA/LIBRA~\cite{DAMA}.

As in QCD, there will presumably be baryons in this model
(corresponding to solitons in the chiral theory, stabilized
by higher derivative terms such as the Skyrme term~\cite{skyrme}), but though
they are SM singlets, they carry DM baryon number, so they
do not couple singly to the SM,  and they do {\em not} look like 
RH neutrinos.


\appendix

\section{\label{sec:app.a}Effects on the Boltzmann equations of the SM particle decays to DMP.}

The decay of the SM particles to the DMP require modification of the Boltzmann equation collision term by adding
two terms $\ccal^h$ and $ \ccal^Z $ corresponding to the $ h \to \pi\pi $ and $ Z \to \pi\pi \pi $ decays.
For the first,
\bea
\ccal^h &=& 2 \int d\Pi_h \, d\Pi_\pi \, d\Pi_\pi 
(2\pi)^4 \delta\up4(p_h - p_\pi - p_\pi)
 |\acal_{h \to \pi + \pi}|^2 f_h (1 + f_\pi) (1 + f_\pi) \cr
&\simeq& N_h\up i m_h \Gamma(h\to \pi\pi)\int \frac{dp^3}{(2 \pi)^3 E_h } f_h \,,  
\eea
where the prefactor of $N_h\up i$ corresponds to the number $\pi_i$ produced, and we approximated $(1 + f_\pi) \simeq 1$. 
Since $\Gamma$ does not depend on $E_h=\sqrt{p^2+m_h^2}$, and using $f_h = e^{-E_h/T}$ (we assume a vanishing Higgs chemical potential), it follows
\bea
\ccal^h &=& N_h\up i m_h \Gamma(h\to \pi\pi) 
\int \frac{dp^3}{(2 \pi)^3 E_h} e^{-E_h/T}
= \frac{N_h\up i}2 \frac{m_h^3}{\pi^2} \frac{K_1(\mx_h)}{\mx_h}  \Gamma(h\to \pi\pi) \cr
&=&  N_h\up i \Gamma(h\to \pi\pi) \, \frac{K_1(\mx_h)}{K_2(\mx_h)} \,\ne_h(\mx_h) \,,
\eea
with $\mx_i$ defined in (\ref{eq:mp-sm.defs}),
 $\Gamma(h \to \pi\pi)$ is given in 
(\ref{eq:hppwidth}),
and where we used (\ref{eq:neq.def}).

In complete analogy, the corresponding contribution from $\Gamma \to \pi \pi \pi$ is 
\beq
\ccal^Z = \Gamma(Z\to \pi\pi\pi)\,\frac{K_1(\mx_Z)}{K_2(\mx_Z)} \,\ne_Z(\mx_Z) \,. 
\eeq
where $\Gamma(Z \to \pi\pi\pi)$ is given in (\ref{eq:Zpppwidth}).
Note that for this decay the final state has a single $ \pi_i $ (and a $ \pi_{\pm\alpbf}$ pair)
so the prefactor corresponding to $N_h\up i$ is $ N_Z\up i=1$.

\section{Kinetics of pure DMP}

Using expressions from Sec.\ref{sec:thermal.history} and Sec.\ref{sec:su2.BE}, and 
Eq.(\ref{eq:p4.scattering}) the Boltzmann equations for pure DMP scattering are
\bea
\frac{dY_i}{d\tau} &=& -
\sum_{j\not=i}\left(Y_i^2 - Y_j^2 \right) 
-\sum_{\alpbf>0} \left (Y_i^2 - Y_\alpbf Y_{-\alpbf} \right) \,,
\cr && \cr
\frac{dY_\alpbf}{d\tau} &=& -
\sum_i \left(Y_\alpbf Y_{-\alpbf} -  Y_i^2 \right) 
- \sum_{\betbf\not=\pm\alpbf,\, \betbf>0}\left(Y_\alpbf Y_{-\alpbf} - 
Y_\betbf Y_{-\betbf} \right)
\,,
\eea
where $ d\tau = \xi\, dx $ with 
\beq
\xi = \sqrt{\frac{\pi g(T)}{45 G}} \frac M{x^2}
\vevof{\sigma v}_{\pi_i\pi_i\to \pi_\alpbf\pi_{-\alpbf}}
\eeq
and the last factor is explicitly given in (\ref{eq:sv.4p}).
We solve these equations in two special cases
\bit
\item Suppose $Y_i = Y_j = Y_C $ for all $i,j$ and $ Y_\alpbf = Y_\betbf
=Y_R$ for all $ \alpbf,\betbf$; then
\beq
\frac{d Y_C}{d\tau}= - \frac{N(N-1)}2 ( Y_C^2 - Y_R^2) \,, \qquad
\frac{dY_R}{d\tau} = - \frac{N-1}2  (Y_R^2 - Y_C^2 ) \,,
\eeq
with solutions
\beq
Y_C = \frac{N^2 \ncal}{N^2-1} \left( w - \inv N \right) \,,\qquad
Y_R = \frac{N^2 \ncal}{N^2-1} \left( 1 - \frac w N \right)\,,
\eeq
where
$ \ncal$ is a constant and
\beq
w = \tanh\left( \frac{N(N-1)}2 \ncal\tau + {\rm const} \right)\,.
\eeq
In particular, $
Y_C(\tau=\infty)=Y_R(\tau=\infty) = N\ncal/(N+1)$.

\item $N=2$. Using the notation of Sec.\ref{sec:su2.BE}
\beq
\frac{dY_o}{d\tau} = Y_+ Y_- - Y_o^2 \,, \qquad
\frac{dY_\pm}{d\tau} = \half ( Y_o^2 - Y_+ Y_-) \,,
\eeq
then\footnote{Other constants of the motion  of the form
$(c_+ + c_-) Y_o + 2\,c_+ Y_+ + 2\,c_- Y_- $ are not independent.}
$ Y_o + Y_+ + Y_- = 3\ncal = {\rm const}$, and $
Y_+ - Y_- = 6 \ncal \delta = {\rm const} $.
Defining now
\beq
\eta=\sqrt{|1-3\delta^2|}\,, \quad
u = \frac{3\eta\ncal}2  \; \tau + u_0\,, \quad
y_{o,\pm} = \frac{Y_{o,\pm}}\ncal \,,
\eeq
where $u_0$ is a constant,  
the time-dependent solutions for $ 3\delta^2 <1 $ are
\bea
y_o = -1 + 2 \eta \;\tanh(u) &\qquad&
y_\pm =  2 \pm 3\delta - \eta \; \tanh(u) \cr
&{\rm or}& \cr
y_o =  -1 + 2 \eta \;\coth(u)  &\qquad&
y_\pm = 2 \pm 3\delta - \eta \; \coth(u) \,,
\eea
where the second set diverges at $ u =0 $;  
in particular, for $\tau\to\infty $: 
$y_o \to -1 +2 \eta $,
$y_\pm\to 2 \pm 3\delta + \eta $
(for $ \tau\to-\infty$ replace $ \eta \to - \eta$).
For $ 3 \delta^2 > 1 $  the 
time-dependent solutions become
\beq
n_o=  -1 - 2 \eta \;\tan(u) \, , \qquad
n_\pm =  2 \pm 3\delta + \eta \; \tan(u) \,,
\eeq
which diverge for $ u = (n + 1/2) \pi,~ n \in \zBB  $.
Note  that for all the time-dependent 
solutions there is always an unphysical  $\tau$ region where $Y_o<0$. 

There are also constant solutions
\bea
y_o&=&  -1 - 2 \eta \qquad 
y_\pm =  2 \pm3 \delta + \eta \,,\cr
y_o&=&  -1 + 2 \eta \qquad 
y_\pm =  2 \pm3 \delta - \eta \,,
\eea
that are real only for $ 3 \delta^2 \le 1 $; note that
the $ \tau$-dependent solutions interpolate between them.
Only the second set has
a region ($|\delta|\le1/2$) where they are all positive,
so these correspond to the steady-state
solutions. 

\eit

\begin{acknowledgments}
The work of SB is supported by U.S Department of Energy under Grant No. DE-SC0008541.
BM acknowledges the support of the Fulbright Foundation and the Ministry of Science and Technology of the Republic
of Croatia under Contract No. 098-0982930-2864. 
\end{acknowledgments}


\bibliographystyle{JHEP}

\end{document}